\documentclass[10pt,journal,compsoc]{IEEEtran}
\usepackage[utf8]{inputenc} 
\usepackage{amsmath}
\usepackage{braket}
\usepackage{graphicx}
\usepackage[]{ragged2e}
\usepackage{pgfplots}
\pgfplotsset{compat=1.15}
\usepackage{csquotes}
\usepackage{enumitem}
\setlist{nolistsep}
\usepackage{lipsum}
\usepackage{hhline}
\def\subparagraph{} % because IEEE classes don't define this, but titlesec assumes it's present
\usepackage{titlesec}
\titlespacing*{\section}{0pt}{*1}{*1}
\titlespacing{\subsection}{0pt}{*1}{*1}

\titlespacing*{\subsubsection}{0pt}{*1}{*1}

\usepackage{algorithm}
\usepackage{algorithmic}

\usepackage{amssymb}

\usepackage{dcolumn}
\usepackage{tabularx}
\usepackage[colorlinks=true,linkcolor=blue,citecolor=blue,urlcolor=blue]{hyperref}
\usepackage{braket}
\usepackage{float}
\usepackage{listings}
\usepackage{color}
\usepackage{subcaption}
\usepackage{placeins}
\usepackage{lipsum}
\usepackage{cuted}
\renewcommand{\thetable}{\Roman{table}}
\newcolumntype{C}{>{\centering\arraybackslash}X}
\newcommand{\RNum}[1]{\uppercase\expandafter{\romannumeral #1\relax}}
\ifCLASSOPTIONcompsoc
  \usepackage[nocompress]{cite}
\else
  \usepackage{cite}
\fi

\ifCLASSINFOpdf
  
\else
  
\fi

\begin{document}
%\title{Securing Image Data in the Quantum Era: An Analysis of Image Encryption with Reduced Quantum Circuit Complexity}
\title{NEQRX: Efficient Quantum Image Encryption with Reduced Circuit Complexity}
%E-NEQR: Efficient Quantum Image Encryption}

\author{Rakesh~Saini,
       Bikash~K.~Behera,
       Saif~Al-Kuwari,
       and~Ahmed~Farouk

%\author{Rakesh~Saini,~\IEEEmembership{Member,~IEEE,}
%         John~Doe,~\IEEEmembership{Fellow,~OSA,}
%         and~Jane~Doe,~\IEEEmembership{Life~Fellow,~IEEE}}% <-this % stops a space
\IEEEcompsocitemizethanks{\IEEEcompsocthanksitem Rakesh Saini is with the Qatar Center for Quantum Computing (QC2), College of Science and Engineering, Hamad Bin Khalifa University, Qatar Foundation, Doha, Qatar.\protect\\
E-mail: rasa68842@hbku.edu.qa
\IEEEcompsocthanksitem Bikash K. Behera is with Bikash's Quantum (OPC) Pvt. Ltd., Balindi, Mohanpur 741246, West Bengal, India.\protect\\
E-mail: bikash@bikashsquantum.com
\IEEEcompsocthanksitem Saif Al-Kuwari is with the Qatar Center for Quantum Computing (QC2), College of Science and Engineering, Hamad Bin Khalifa University, Qatar Foundation, Doha, Qatar.\protect\\
E-mail: smalkuwari@hbku.edu.qa
\IEEEcompsocthanksitem Ahmed Farouk is with the Department of Computer Science, Faculty of Computers and Artificial Intelligence, South Valley University, Hurghada, Egypt.\protect\\
E-mail: ahmed.farouk@sci.svu.edu.eg
}
}
\IEEEtitleabstractindextext{%
\justify\begin{abstract}
Cryptography plays an important role in ensuring data security and authentication within information processing systems. As the prevalence of digital imagery continues to grow, safeguarding this form of data becomes increasingly crucial. However, existing security protocols, reliant on complex mathematical models, exhibit vulnerabilities in effectively protecting information from both internal and external threats. Moreover, the forthcoming advent of quantum computing poses a significant challenge, as it could decrypt data encrypted by classical. %cryptographic algorithms, %jeopardizing the security of private data.Compounding this challenge is the impending era of quantum computing, which threatens the efficacy of classical cryptographic algorithms, 
%thereby compromising the security of sensitive data.
In this paper, we propose an efficient implementation scheme for a quantum image encryption algorithm combining the generalized affine transform and logistic map.
%on IBM quantum computers. 
%Our approach capitalizes on generalized affine transform and logistic map methodologies, harnessing the intrinsic power of quantum computing to fortify encryption protocols. 
%We rigorously evaluate our solution by implementing quantum circuits using Qiskit and quantum devices, substantiating the efficacy and robustness of our encryption technique across diverse performance metrics.
We evaluated developed quantum circuits using qiskit and quantum devices to validate the encryption technique. Through comprehensive performance analysis, we have demonstrated the efficiency of the chosen encryption algorithm across various criteria.
Furthermore, we introduce a hybrid methodology aimed at mitigating circuit complexity and reducing quantum cost. Leveraging the Espresso algorithm and incorporating an ancilla qubit into the circuitry, we achieve a remarkable 50\% reduction in cost while maintaining security and efficiency.
Finally, we conducted robustness and security analyses to assess the resilience of our encryption method against diverse noise attacks. The results confirm that our proposed quantum image encryption technique provides a secure solution and offers precise and measurable quantum image processing capabilities.
\end{abstract}

}

\maketitle

\IEEEdisplaynontitleabstractindextext

\IEEEpeerreviewmaketitle

\ifCLASSOPTIONcompsoc
\IEEEraisesectionheading{\section{Introduction}\label{sec:introduction}}
\else
\section{Introduction}
\label{sec:introduction}
\fi

\IEEEPARstart{I}{n} the landscape of information science, digital image processing has emerged as a fundamental discipline since its inception in the 1960s \cite{NEQR_Gonzalez2002}.  Imaging processing plays an essential role in various fields such as remote sensing, robot vision, pattern recognition, medical field, and many other areas. At its core, digital image processing involves the analysis, transmission, and manipulation of digital imagery, which are represented as two-dimensional arrays or matrices of numbers obtained through sampling and quantization of analog images.
%It is the summation of analyzing, transmitting, and manipulating digital images. Digital images are a two-dimensional array or matrix of numbers obtained by performing sampling and quantization of analog images. 
The exponential growth in digital image acquisition and sharing and the widespread implementation of machine learning, the Internet of Things, and other technologies have increased the demand for the quality and security of digital images. The quality of digital images depends on factors such as resolution and pixel count, while security concerns revolve around preventing eavesdroppers from accessing sensitive image information. Addressing these tasks requires robust cryptography techniques capable of withstanding sophisticated attacks while also demanding substantial computational and storage resources.
%higher computation, and storage capacity. 
However, existing computing models are not able to meet all these requirements. 

Currently, the classical cryptosystems contain data encryption standards (DES), triple-DES, advanced encryption standards, elliptic curve cryptography, and Rivest Shamir Adelman, to name a few. However, while these algorithms are excellent for encryption, they remain vulnerable to the emerging threat posed by quantum computers \cite{NEQR_FeynmanIJTP1982}.
Quantum computers have already demonstrated the ability to perform real-time factorization using quantum algorithms, thereby undermining the security of classical cryptographic methods.
%Quantum computers have already demonstrated that using quantum algorithms factorization in real-time is feasible, which is the base of the security of most classical cryptographic methods.  
Moreover, as quantum computers continuously prove their superiority over classical methods, it is pretty close that they will become a reality.
Therefore, to ensure the confidentiality and integrity of digital imagery in quantum computing era, it has become necessary to develop encryption techniques using quantum computing principles and achieve post-quantum cryptography \cite{NEQR_Gisin2002}. 
%Therefore, to solve such problems, Feynman in 1982 introduced a different computing model based on quantum mechanical laws, mainly quantum superposition
%and quantum entanglement \cite{NEQR_FeynmanIJTP1982}. The main advantage of
%using a quantum computing model in information science is that it provides massive parallel computation \cite{NEQR_DeutschRSLA1985}, and unconditional security \cite{NEQR_Gisin2002} to the data. That is why the quantum computing model is vastly used in information science.

Various proposals have emerged to compute and store the digital image on a quantum computer \cite{NEQR_Yan2016, NEQR_Su2020}, such as qubit lattice \cite{NEQR_VenegasSPIECQIC2003}, real ket \cite{NEQR_Latorre2005}, entangle image \cite{NEQR_Andraca2010}, flexible representation of quantum images (FRQI) \cite{NQER_Le2011}, multi-channel representation for images (MCQI) \cite{NEQR_Sun2013}, novel enhanced quantum representation (NEQR) \cite{NEQR_Zhang2013a}, novel quantum representation of color digital images (NCQI) \cite{NEQR_Sang2017}, A Generalized Model of NEQR (GNEQR) \cite{NEQR_Li2019}, normal arbitrary quantum superposition state (NAQSS) \cite{NEQR_Li2014}, and quantum Boolean image processing (QuBoIP) \cite{NEQR_Mastriani2014b}. Of particular interest is the NEQR method, which is very similar to the FRQI except for its approach to encoding the pixel values. NEQR uses eight qubits, hence $2^8$ basis states to encode 256 different pixel values, while FRQI uses different polar and azimuthal angles of a single qubit state. Therefore, NEQR exhibits higher measurement accuracy than FRQI. In addition, quantum encryption algorithms have been developed to transform the original image into a cipher image to secure the information in the image. These encryption procedures include scrambling the pixel positions and/or altering the pixel values using chaos theory or other transformations. These encryption procedures can be categorized into two main types: 1) transforming the image into a frequency domain with random operations \cite{NEQR_LiangQIP2016, NEQR_Jiang2014, NEQR_Yang2014} and 2) converting the image into an encrypted image using chaos theory \cite{NEQR_Devaney2003,NEQR_LiangQIP2016,NEQR_RAn2018,NEQR_Zhou2018}. In this study, we focus on the latter type of encryption algorithm by including a logistic map to encrypt the image parameters. 

In this paper, we propose an efficient implementation scheme for a %propose a 
novel quantum encryption technique that combines Novel Enhanced Quantum Representation (NEQR) (encoding procedure) \cite{NEQR_Zhang2013a}, generalized affine transform (scrambles the pixel position) \cite{NEQR_LiangQIP2016}, and logistic map (changes pixel value using chaos theory) \cite{NEQR_LiangQIP2016}. The quantum circuits and their corresponding algorithms were executed separately on an IBM quantum computer. Then, the Espresso algorithm \cite{NEQR_Brayton}, used in \cite{NEQR_Zhang2013a}, which optimizes circuit complexity has been used with our modified approach. Moreover, the analysis of keyspace, encryption procedure quality, histogram, circuit complexity, and variation in the fidelity of the NEQR circuit is performed under different effectiveness of the noisy channels, including amplitude-damping, phase-damping, bit-flip, phase-flip, bit-phase-flip, and depolarizing.
%%%%%%%%%%%%%%%% until here, you need to revamp

%this is generally OK
The contributions of this article can be summarized as follows:
\begin{enumerate}
    \item We propose an efficient scheme for quantum image encryption based on GAT and LM. % on IBM quantum computer. 
    \item We introduce a hybrid approach to reduce circuit complexity and quantum cost using the Espresso algorithm, which reduced the cost by approximately 50\%.% for the cost of adding one more qubit to the circuit.
    %\item We produced a prototype implementation of the proposed algorithm and simulated it on the IBM Quantum cloud. 
    %\item The proposed algorithms' performance, robustness, security, and computational complexity analysis are measured for their efficiency against various criteria and noise attacks.
\end{enumerate}

%This is generally OK
The rest of this paper is organized as follows: Section \ref{sec2} briefly reviews some preliminaries. Section \ref{sec3} presents our proposed quantum image encryption algorithm. The evaluation of the keyspace, quality of encryption procedure, histogram, circuit complexity, and robustness test of NEQR circuit in different noisy channels are presented in Section \ref{sec4}. Finally, Section \ref{sec5} concludes the paper.
%and proposes future extensions to this work.

\section{Preliminaries}\label{sec2}
\subsection{Novel Enhanced Quantum Representation (NEQR)}
NEQR, introduced by Zhang et al. \cite{NEQR_Zhang2013a}, offers a novel approach to quantum gray image representation. This model uses $2^n \times 2^n$ gray image parameters as input and encodes them into a quantum circuit. The construction of this circuit involves $2n + 8$ qubits, with 8 qubits designated for storing the gray-scale value $f(Y, X)$ of the pixels, while the remaining $2n$ qubits represent the pixel positions (Y, X). The resulting state of this quantum circuit encapsulates the entirety of the image. Mathematically, the final state can be expressed as follows:
\begin{align}
    \Ket{\psi}= \frac{1}{2^n}\sum_{Y, X=0}^{2^{2n} - 1} \Ket{f(Y, X)}\Ket{YX},
\end{align}
where $\Ket{f(Y, X)}$ can be written in the bit sequence as, $\bigotimes_{i = 0}^{p - 1}\Ket{C_{YX}^{i}} =  \Ket{c^{7}_{YX}c^{6}_{YX}\dots c^{1}_{YX}c^{0}_{YX}}$, which represents the encoded gray-scale value of the pixel, and 
\begin{align}
    \Ket{YX} &= \Ket{Y}\Ket{X} \nonumber\\&= \Ket{Y_{n-1}Y_{n-2}\dots Y_{1}Y_{0}}\Ket{X_{n-1}X_{n-2}\dots X_{1}X_{0}},\nonumber
\end{align}
is used to store the corresponding vertical and horizontal coordinate values of the pixels.
%-------------------------
% NEQR Algorithm
%------------------------------------
\begin{algorithm}[]
\caption{Algorithm to encode digital images into a quantum circuit using NEQR.}
\label{NEQR_algo1}
\begin{algorithmic}[1]
\renewcommand{\algorithmicrequire}{\textbf{Input:}}
\renewcommand{\algorithmicensure}{\textbf{Output:}}
\REQUIRE Classical image: $2^n \times 2^n$\\
Declare the Pixel value: $f(Y, X)$\\
Declare the vertical and horizontal location \\of pixels: $(Y, X)$\\
Quantum Circuit: QC \{initial state $\Ket{0}^{2n+8}$\}\\
\ENSURE  Quantum Circuit for given classical image
\\ \textit{Initialisation} :
\STATE Start by transforming the position and pixel value in binary sequence: \\
$(Y, X)$ $\rightarrow$ $\ket{YX}$ and $f(Y, X) \rightarrow \bigotimes_{i = 0}^{7}\Ket{C_{YX}^{i}}$\\
\STATE To store these binary numbers in the quantum circuit follow:\\
\textbf{(I)} The initial state of the circuit: $\ket{0}^{2n + 8}$
\\
\textbf{(II)} Apply Hadamard gate on $2n$-qubits that generate superposition of $2^{2n}$ states:
\\
$\Ket{0}^8 \otimes(H^{2n}\otimes\Ket{0}^{2n}) \rightarrow \frac{1}{2^{n}}\sum_{Y = 0}^{2^{n} - 1}\sum_{X = 0}^{2^{n} - 1} \Ket{0}^8\otimes\Ket{YX}$
\\
\textbf{(III)} Define an operation $O_{YX}$ that apply a unitary operation $U_{YX}$ on the pixel value qubits for pixel state $\Ket{YX}$ as:\\
% \vspace{.15cm}
% \linebreak
$O_{YX} = I \otimes \sum_{j = 0}^{2^{2n} - 1}\sum_{i = 0, ij\neq YX}^{2^{2n} - 1} \ket{ij}\bra{ij} + U_{YX} \otimes \ket{YX}\bra{YX}$\\
% \vspace{.15cm}
% \linebreak
$U_{YX}\Ket{0} \rightarrow \Ket{0 \oplus C_{YX}}$
\\
\textbf{(IV)} Now, operate $O_{YX}$ on all qubits:\\
\vspace{.15cm}
% \linebreak
$\frac{1}{2^{n}}O_{YX}\left(\sum_{Y = 0}^{2^{n} - 1}\sum_{X = 0}^{2^{n} - 1} \Ket{0}^8\otimes\Ket{YX}\right)\rightarrow$\\
\vspace{.15cm}
% \linebreak
$\frac{1}{2^{n}}\sum_{Y = 0}^{2^{n} - 1}\sum_{X = 0}^{2^{n} - 1} (U_{YX}\otimes\Ket{0}^8)\otimes\Ket{YX}\rightarrow$\\
\vspace{.15cm}
% \linebreak
$\frac{1}{2^{n}}\sum_{Y = 0}^{2^{n} - 1}\sum_{X = 0}^{2^{n} - 1} \Ket{f(Y, X)}\otimes\Ket{YX}\rightarrow\Ket{\psi}$\\
\STATE\textbf{return} QC, $\Ket{\psi}$
\end{algorithmic}
\end{algorithm}

\begin{figure}
    \centering
    \includegraphics[width=0.4\textwidth]{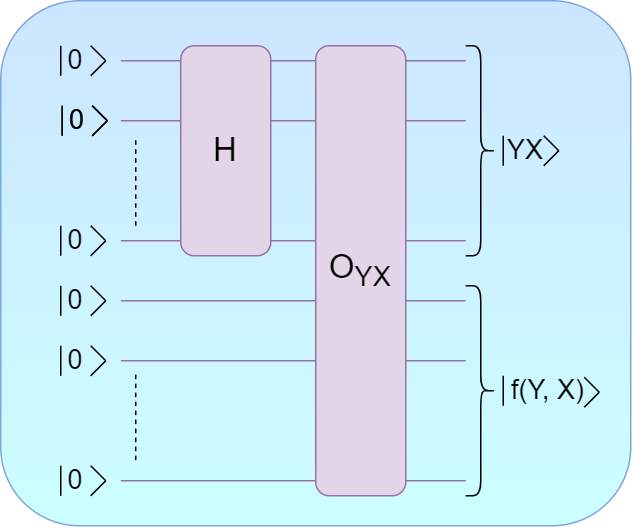}
    \caption{NEQR procedure circuit using $(2n + 8)$-qubits. Here, H is the Hadamard gate and $O_{YX}$ is a controlled gate that operate $U_{YX}$ on pixel qubits, when 2n-qubits are in the state $\Ket{YX}$, to obtain $f(Y, X)$.}\label{NEQR_fig1}
\end{figure}

For example, upon performing NEQR on an image of size $2 \times 2$ as shown in Fig. \ref{NEQR_fig8a}, it can be represented by the quantum state:
\begin{align}
    \Ket{\psi_I}&=\frac{1}{2}(\Ket{11111111}\Ket{00}+\Ket{00000000}\Ket{01}\nonumber\\
    &\quad +\Ket{11001000}\Ket{10}+\Ket{01100100}\Ket{11}).
    \label{image_eq}
\end{align}
Algo. \{\ref{NEQR_algo1}\} describes the NEQR procedure for encoding digital images into a quantum circuit, and Fig. \ref{NEQR_fig1} shows the general quantum circuit for the NEQR method.

\subsection{Generalized Affine Transform}
\begin{figure}[]
\centering
\includegraphics[width = 0.5\textwidth]{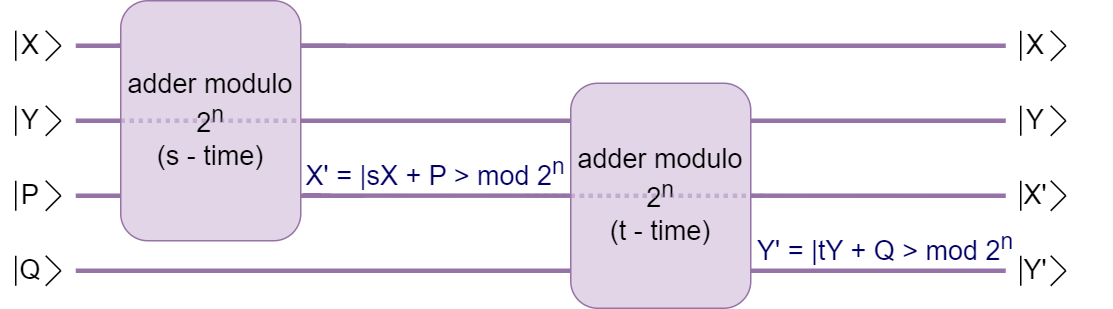}
\caption{A quantum circuit for GAT to scramble the original pixel coordinate $\ket{YX}$ into $\Ket{Y'X'}$. The dotted qubit line in the circuit represents bypassing the operation.}\label{NEQR_fig2}
\end{figure}

\begin{figure*}[]
    \centering
    \includegraphics[width = \textwidth]{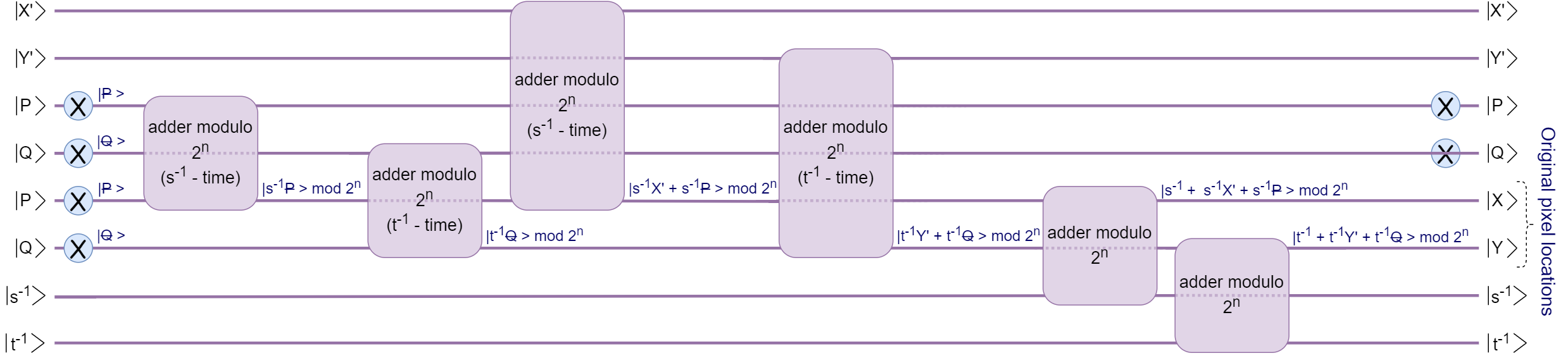}
    \caption{Quantum circuit represents a General decryption procedure for GAT to obtain the real pixel coordinate $\ket{YX}$ from $\Ket{Y'X'}$. The doted qubit line refers to avoiding the applied operation.}\label{NEQR_fig4}
\end{figure*}

GAT, shown in Algo. \{\ref{NEQR_algo2}\}, scrambles the pixel locations $\Ket{YX}$ of the image. Here, $X$ and $Y$ represent the actual horizontal and vertical locations of pixels, respectively \cite{NEQR_LiangQIP2016}. The mathematical representation of GAT, as described in, is given by:
\begin{align}
    \begin{pmatrix}
        X' \\
        Y'
    \end{pmatrix}
    =
    \begin{pmatrix}
        s & 0 \\
        0 & t
    \end{pmatrix}
    \begin{pmatrix}
        X \\
        Y
    \end{pmatrix}
        +
    \begin{pmatrix}
        P \\
        Q
    \end{pmatrix}
    \text{mod $2^n$},
    \label{Af}
\end{align}
where $X'$ and $Y'$ represent the scrambled horizontal and vertical locations of pixels, taking the form $(sX + P)$ and $(tY + Q)$, respectively, with modulo $2^n$. The parameters $s$, $t$, $P$ \& $Q$ are for GAT and obey the following conditions: 
\begin{itemize}
     \item $P$ and $Q$ should not be zero. In the circuit implementation, $n$ qubits are allocated for each to store their values.
     \item $s$ and $t$ should be co-prime with $2^n$.
\end{itemize}
The generalized circuit in Fig. \ref{NEQR_fig2} illustrates the operations involved in scrambling the original coordinates $\ket{X}$, $\ket{Y}$ to $\ket{X'}$, $\ket{Y'}$, as calculated in Equation (\ref{Af}).
The inverse operation of GAT, depicted in Algo. {\ref{NEQR_algo4}}, is utilized to recover the original pixel coordinates from the scrambled ones. Mathematically, this inverse operation can be expressed as: 
\begin{align}
     \begin{pmatrix}
        X \\
        Y
        \end{pmatrix}
        =
        \begin{pmatrix}
        s^{-1} & 0 \\
        0 & t^{-1}
    \end{pmatrix}
    \begin{pmatrix}
        X' - P \\
        Y' - Q
    \end{pmatrix}
         \text{mod $2^n$}.
         \label{inv_AF_eq}
\end{align}
This equation gives the original coordinates of the pixel $X = s^{-1}(X' - P)$ $mod$ $2^{n}$ and $Y = t^{-1}(Y' - Q)$ $mod$ $2^{n}$. Here, $s^{-1}$ and $t^{-1}$ are the modular multiplicative inverse \cite{NEQR_Rosen} of $s$ and $t$, respectively, and can be calculated as follows:
\begin{align}
      1 &= ss^{-1} \text{mod $2^n$},\nonumber\\
      1 &= tt^{-1} \text{mod $2^n$}.
\end{align}
The generalized circuit in Fig. \ref{NEQR_fig4} shows the operations performed to recover the original coordinates $\ket{X}$, $\ket{Y}$ from $\ket{X'}$, $\ket{Y'}$ as calculated in Eq. (\ref{inv_AF_eq}).
%------------------------------------
%generalized affine Transformation Encryption Procedure
%------------------------------------
\begin{algorithm}[]
\caption{Algorithm to encrypt pixel positions by using GAT.}
\label{NEQR_algo2}
\begin{algorithmic}[1]
\renewcommand{\algorithmicrequire}{\textbf{Input:}}
 \renewcommand{\algorithmicensure}{\textbf{Output:}}
 \REQUIRE Declare the size of the image: $2^n \times 2^n$\\
Declare the GAT parameters: $P$, \\$Q$, $s$ and $t$\\
Declare the horizontal and vertical location \\of pixels: $X$, $Y$\\
 \ENSURE  Image of scrambled pixel location
\\ \textit{Initialisation} :
\STATE $i$, $j \gets 1$
\WHILE {$i \leq s$}
    \STATE $X' = (iX + P)$ mod $2^n$
    \STATE $i \gets i + 1$
\ENDWHILE
\WHILE{$j \leq t$}
    \STATE $Y' = (jY +Q)$ mod $2^n$
    \STATE $j \gets j + 1$
\ENDWHILE  
\STATE \textbf{return} $X'$, $Y'$
\end{algorithmic}
 \end{algorithm}

\subsection{Adder modulo $2^{n}$}
\begin{figure}[]
    \centering
     \begin{subfigure}[]{0.5\textwidth}
        \begin{minipage}{0.02\linewidth}
        \centering
        \caption{}\label{NEQR_fig3a}
        \end{minipage}\hfill
        \begin{minipage}{.96\linewidth}
        \centering
        \includegraphics[scale=0.3]{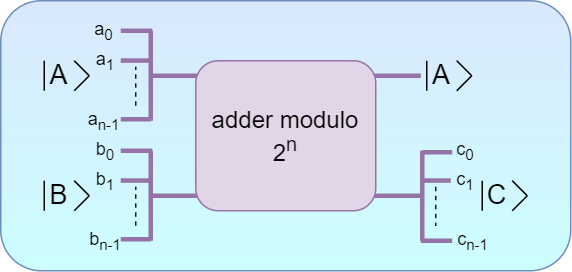}
        \end{minipage}
     \end{subfigure}
     \begin{subfigure}[]{0.55\textwidth}
     \vspace*{20pt}
        \begin{minipage}{0.02\linewidth}
        \centering
        \caption{}\label{NEQR_fig3b}
        \end{minipage}\hfill
        \begin{minipage}{.96\linewidth}
        \centering
        \includegraphics[scale=0.3]{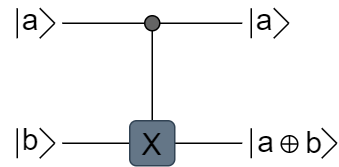}
        \end{minipage}
     \end{subfigure}
    \caption{(a) Adder modulo circuit: with output $\Ket{C} = \Ket{A} \oplus \Ket{B}$ mod $2^n$. (b) Quantum CNOT gate: gives output as $a$ $\oplus$ $b$ $mod$ $2$.}\label{NEQR_fig3}
\end{figure}
The adder modulo $2^n$ is designed to sum two binary numbers, each represented by $n$-bits, ensuring that the resultant binary number maintains the same number of bits. This is achieved by ignoring the last carry bit. A generalized quantum circuit for the adder modulo $2^n$ is given in \cite{NEQR_JiangQIP2014-2}. A general black-box representation of this circuit for $2n$-bits input is shown in Fig. \ref{NEQR_fig3a}. For $n = 1$, the adder modulo $2^n$ circuit simplifies to the quantum CNOT gate, shown in Fig. \ref{NEQR_fig3b}.

\subsection{Logistic Map}\label{Logistic_Map}
LM is used to iteratively map the value of the previous step to the next step \cite{NEQR_Devaney2003}, using the following dynamic equation:
\begin{align}
    L_{\eta + 1} = \delta L_{\eta}(1-L_{\eta}),
    \label{L_eq}
\end{align}
where $\eta = 0, 1, 2, ...., 2^{2n}-1$. Here, $L_{0} \epsilon {0,1}$ denotes the initial value of LM, $\delta$ represents the growth rate, and $2^{2n}$ signifies the total number of pixels.

This dynamic equation exhibits chaotic behavior for 3.85 $\leq\delta\leq$ 4, which means that with this growth rate, Eq. (\ref{L_eq}) generates pseudo-random strings of 0's and 1's.

%%%%%%%%%%%%%%%%%%%%%%%%%%%%%%%%%%%%%%%%%%%%%%%%%%%%%%%%%%%%%%%%%
\section{Proposed Algorithm}\label{sec3}
\subsection{NEQR Circuit}
In Fig.\ref{NEQR_fig10a}, where $n = 1$, a quantum circuit of $2+8$-qubits is initialized in all 0's state. The first $2$-qubits store the pixel coordinates, while the rest represent pixel values. The encoding procedure of the image parameters in a quantum circuit, as outlined in Algorithm \{\ref{NEQR_algo1}\}, is as follows:
\begin{enumerate}
    \item The image, shown in Fig. \ref{NEQR_fig8a}, has $4$-pixels with different positions and values. To process all pixel positions, the Hadamard gates on the first $2$-qubits are employed that generate a superposition of four equally probable states $(\Ket{00}$, $\Ket{01}$, $\Ket{10}$, and $\Ket{11})$.
    \item Toffoli gates are then applied, using coordinate qubits as control and the remaining qubits as targets. These Toffoli gates transform the pixel value qubits from state $\bigotimes_{i = 0}^{7}\Ket{0^{i}}$ to state $\bigotimes_{i = 0}^{7}\Ket{C_{YX}^{i}}$, corresponding to each pixel $\Ket{YX}$.
\end{enumerate}
The conversion of the state of the quantum circuit is as follows:
\begin{align}
      \ket{0000000000}&\xrightarrow[\text{first $2$-qubits}]{\text{apply H-gate on }} \Ket{00000000}\ket{++}\nonumber\\
      &\xrightarrow[ \text{on rest qubits}]{\text{apply Toffoli gates}} \Ket{\psi_{I}}.
\end{align}
The resultant quantum circuit for this whole procedure is shown in Fig. \ref{NEQR_fig10a}, featuring $14$ Toffoli gates. However, before executing on real superconducting devices, the circuit must be transpiled to the device's basis gates, which can exponentially increase circuit size and depth, impacting execution efficiency and fidelity.

\begin{figure}
    \centering\includegraphics[width=0.45\textwidth]{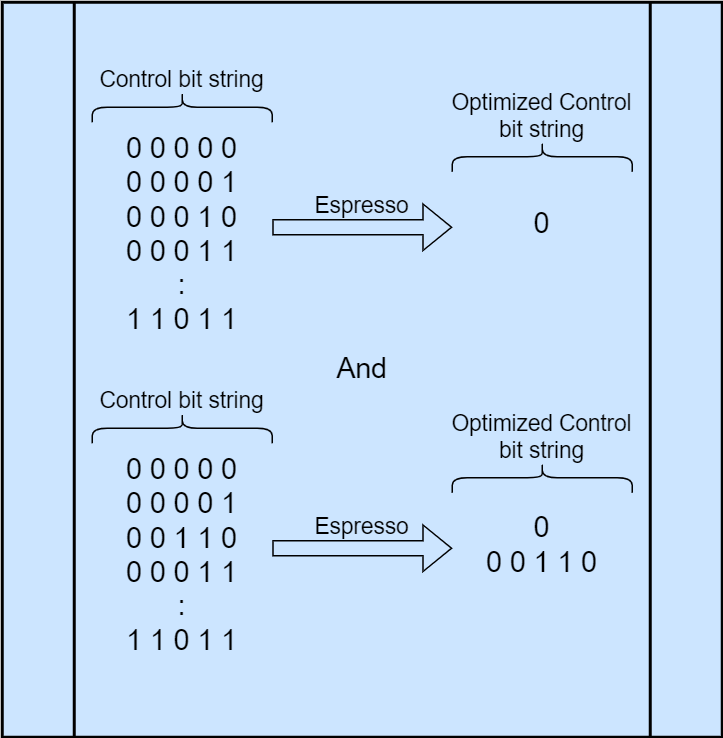}
    \caption{An example to demonstrate the optimization of control bit strings using the Espresso algorithm. The original control bits strings on the left and the optimized control bit strings on the right can perform the same operation on the $i^{th}$ target qubit.}\label{NEQR_5}
\end{figure}
To overcome this problem, the image compression in \cite{NEQR_Zhang2013a} is used. This procedure handles the control bit strings of the pixel values using the Espresso algorithm \cite{NEQR_Brayton} so that a similar task can be performed with a lower control bit string, as shown in Fig. \ref{NEQR_5}. As a result, the circuit complexity is reduced. In this compression procedure, an extra step has been added to reduce the circuit complexity further. This step uses an ancillary qubit to store the information of the Toffoli gates, which are then used as control and target the desired qubits one by one using C-NOT gates. The same task can be implemented using a few Toffoli gates, C-NOT gates, and one ancillary qubit, see Fig. \ref{NEQR_fig10c}. This step is used when the number of Toffoli gates with the same control bit string is still high after using the image compression procedure \cite{NEQR_Zhang2013a}. Fig. \ref{NEQR_fig10b} and Fig. \ref{NEQR_fig10d} illustrate a comparison.
\subsection{Encryption procedure}
The encryption procedure utilizes a combination of LM and GAT to transform the original image into an encrypted form. %Here, the algorithms for encryption schemes and how they affect image parameters are discussed and presented. 
\subsubsection{logistic map (Diffusion stage)}
The diffusion stage is applied to the value of the pixel and is controlled by the corresponding pixel coordinates. This stage utilizes chaos theory as described in Section \ref{Logistic_Map}, resulting in a chaotic image. The quantities required to operate LM in a quantum circuit are calculated according to Algo. \{\ref{NEQR_algo3}\}. The operations of LM in a quantum circuit are as follows:
\begin{enumerate}
    \item Pick the output lists $J$, $T$ of Algo. \{\ref{NEQR_algo3}\}.
    \item Transform each $J_{YX}$ and $T_{YX}$ form $J$ and $T$ to a binary sequence as:
    \begin{align}
        \Ket{J_{YX}} = \otimes_{i=0}^{p-1}\Ket{J_{YX}^i},\quad \Ket{T_{YX}} = \otimes_{i=0}^{p-1}\Ket{T_{YX}^i}.
    \end{align}
    \item Defining quantum operations $D_{YX}$ and $E_{YX}$ to manipulate pixel values based on $J_{YX}$ and $T_{YX}$ as:
    \begin{align}
          D_{YX}\Ket{C_{YX}^i}& \rightarrow \Ket{C_{YX}^i\oplus J_{YX}^i},\nonumber\\
          E_{YX}\Ket{C_{YX}^i \oplus J_{YX}^i}& \rightarrow \Ket{C_{YX}^i\oplus J_{YX}^i \oplus T_{YX}^i}.\nonumber
    \end{align}
    \item Define functions that apply $D_{YX}$ and $E_{YX}$, 
    \begin{align}
          \phi_{YX}^1& = I \otimes \frac{1}{2^n}\sum_{j=0}^{2^{n} - 1}\sum_{i = 0, ji\neq YX}^{2^{n} - 1}\Ket{ji}\bra{ji} + D_{YX} \nonumber\\ &\quad\otimes \Ket{YX}\bra{YX},\nonumber\\
          \phi_{YX}^2& = I \otimes \frac{1}{2^n}\sum_{j=0}^{2^{n} - 1}\sum_{i = 0, ji\neq YX}^{2^{n} - 1}\Ket{ji}\bra{ji} + E_{YX} \nonumber\\ &\quad\otimes \Ket{YX}\bra{YX}.\nonumber
    \end{align}
    \item Apply $\phi_{YX}^1$ and $\phi_{YX}^2$ on the quantum state $\psi_{I}$ as:
    \begin{align}
          \Ket{\psi_{L}} &= \phi_{YX}^{2}\phi_{YX}^1\Ket{\psi_I}\nonumber\\
        &= \frac{1}{2^n}\sum_{Y, X=0}^{2^{2n} - 1} (E_{YX}D_{YX}\bigotimes_{i = 0}^{p - 1}\Ket{C_{YX}^{i}})\Ket{YX}\nonumber\\
        &= \frac{1}{2^n}\sum_{Y, X=0}^{2^{2n} - 1} \bigotimes_{i = 0}^{p - 1}\Ket{C_{YX}^i\oplus J_{YX}^i \oplus T_{YX}^i}\Ket{YX}.\nonumber
    \end{align}
\end{enumerate}
\subsubsection{Generalized Affine Transform (Permutation stage)}
GAT, represented by the operator $A$, operates only on the pixel positions of the quantum state of the image $\Ket{\psi_L}$. %let $A$ represent the operator for the generalized affine transform, which only affects the pixel positions of the image.  When A operates on the quantum state of the image $\Ket{\psi_L}$, it only focuses on the position of the pixels $\Ket{YX}$. 
The operation is as follows:
\begin{align}
    \Ket{\psi_{AL}} = A\Ket{\psi_L} &= \frac{1}{2^n}\sum_{Y = 0}^{2^{n} - 1}\sum_{X = 0}^{2^{n} - 1} \Ket{f'(Y, X)}A\Ket{YX}\nonumber\\
    &= \frac{1}{2^n}\sum_{Y = 0}^{2^{n} - 1}\sum_{X = 0}^{2^{n} - 1} \Ket{f'(Y, X)}\Ket{Y'X'},\nonumber
\end{align}
where $A$ scrambles the horizontal and vertical positions of the pixels as $A\Ket{X} = \Ket{X'}$ and $A\Ket{Y} = \Ket{Y'}$, respectively. This transformation is described in Algo. \{\ref{NEQR_algo2}\} and the output of this algorithm $\Ket{X'}$ and $\Ket{Y'}$ are given by:
\begin{align}
      \Ket{X'}& = \Ket{sX + P} \text{mod $2^n$},\nonumber\\
      \Ket{Y'}& = \Ket{tY + Q} \text{mod $2^n$}.\nonumber
\end{align}
According to the quantum version of GAT to scramble pixel positions, a general quantum circuit given in Fig. \ref{NEQR_fig2} is designed. This quantum circuit performs the adder modulo $2^n$ operation, $s$ and $t$ times to transform the states $\Ket{X} \rightarrow \Ket{X'}$ and $\Ket{Y} \rightarrow \Ket{Y'}$, respectively, as shown in Fig. \ref{NEQR_fig2}. The algebra of quantum circuit is as follows:
\begin{align}
      \Ket{X, P} &\xrightarrow[\text{s - times}]{\text{adder modulo $2^n$}} \Ket{X, (sX + P) \text{mod $2^n$}},\nonumber\\
      \Ket{Y, Q} &\xrightarrow[\text{t - times}]{\text{adder modulo $2^n$}} \Ket{Y, (tY + Q) \text{mod $2^n$}}.\nonumber
\end{align}
%-----------------------------
%Logistic map encryption procedure
%-----------------------------
\begin{algorithm}[]
\caption{Algorithm to encrypt pixel values by using LM.}
\label{NEQR_algo3}
\begin{algorithmic}[1]
\renewcommand{\algorithmicrequire}{\textbf{Input:}}
\renewcommand{\algorithmicensure}{\textbf{Output:}}
\REQUIRE Declare the size of the image: $2^n \times 2^n$\\
Declare the LM parameters: $\delta$, $L_0$\\
\ENSURE  $J_{YX}$, $T_{YX}$ \{$YX = \eta$\}
\\ \textit{Initialisation} :
\STATE $J$, $T = [\quad]$
\FORALL{$\eta$ in range$(2^{2n}-1)$}
    \STATE $L_{\eta + 1} = \delta L_{\eta} (1 - L_{\eta})$
    \STATE $J_{YX} = round(mod(L_\eta \times 2^8, 2^8))$
    \STATE $J.append(J_{YX})$
\ENDFOR
\FORALL{$\eta$ in range$(2^{2n}-1)$}
    \STATE $T_{YX} = J_{2^{2n} - 1 - YX}$
    \STATE $T.append(T_{YX})$
\ENDFOR
\STATE \textbf{return} $J$, $T$
\end{algorithmic}
\end{algorithm}
\subsection{Decryption procedure}
Decryption involves reversing GAT and LM procedures.
\subsubsection{Inverse Generalized Affine Transform}
The inverse GAT retrieves the original horizontal and vertical pixel coordinates of the image. Let $A^{-1}$ be the inverse GAT operation:
\begin{align}
      \Ket{\psi_L} &= A^{-1}\Ket{\psi_{AL}}\nonumber\\ &= \frac{1}{2^n}\sum_{Y, X=0}^{2^{2n} - 1} \Ket{f'(Y, X)}A^{-1}\Ket{Y'X'}\nonumber\\
     &= \frac{1}{2^n}\sum_{Y, X=0}^{2^{2n} - 1} \Ket{f'(Y, X)}\Ket{YX}.\nonumber
\end{align}
According to Theorem 3 in \cite{NEQR_Jiang2014}, the negative adder modulo $2^n$ in the Eq. (\ref{inv_AF_eq}) can be realized as:
\begin{align}
    (x - y) \text{$mod$ $2^n$} = (x + (\overline{y} + 1)) \text{$mod$ $2^n$},\nonumber
\end{align}
where $\overline{y}$ = $\overline{y}_{n-1} \overline{y}_{n-2}\dots\overline{y}_{0}$ = 1 - $y_i$ with $i = n-1, n-2,\dots,0$. Thus, $X = s^{-1}(X' - P)$ $mod$ $2^n$ is converted to $X = s^{-1}(X' + (\overline{P} + 1))$ and similarly for $Y$. The circuit in Fig. \ref{NEQR_fig4} takes $2s^{-1} + 3$ steps to obtain the original pixel position $\Ket{X}$ and $2t^{-1} + 3$ for $\Ket{Y}$. Refer to Algo. \{\ref{NEQR_algo4}\} for a  step-by-step explanation of the inverse GAT procedure.
%-----------------------------
%affine Transformations Decryption Procedure
%-----------------------------
\begin{algorithm}[]
\caption{Algorithm to decrypt the pixel values using GAT.}
\label{NEQR_algo4}
\begin{algorithmic}[1]
 \renewcommand{\algorithmicrequire}{\textbf{Input:}}
 \renewcommand{\algorithmicensure}{\textbf{Output:}}
 \REQUIRE Declare the GAT parameters: $P$, $Q$, $s$ and $t$\\
Declare the scrambled image pixel location:\\ $X'$, $Y'$ \\
Quantum Circuit for the location of pixels: $QC_1$\\
($QC_1$ has $2n$ qubits)\\
Quantum Circuit to store $P$ and $Q$: $QC_2$\\
($QC_2$ has $2n$-qubits, $n$ to store $P$ and rest for \\$Q$)\\
GAT operator: $A$ (Adder modulo $2^n$)
 \ENSURE  Original image
 \\ \textit{Initialisation} :
\STATE Apply $A$ on $\overline{P}$ \{as control\} to $\overline{P}$ \{as target\} 
\FORALL{$i$ in range$(1, 2s^{-1} + 4)$}
    \FORALL{$1 \leq i \leq s^{-1}$}
    \STATE $\overline{P} \gets (i\overline{P})$ mod $2^n$
    \ENDFOR
    \FOR{$i = s^{-1} + 1$}
    \STATE Replace $\overline{P}\{control\}$ to $X'\{control\}$
    \ENDFOR
    \FORALL{$s^{-1} +2 \leq i \leq 2s^{-1} + 1$}
    \STATE $\overline{P} \gets (iX' + s^{-1}\overline{P})$ $mod$ $2^n$
    \ENDFOR
    \FOR{$i = 2s^{-1} + 2$}
    \STATE Replace $X'\{control\}$ to $s^{-1}\{control\}$
    \ENDFOR
    \FOR{$i = 2s^{-1} + 3$}
    \STATE $\overline{P} \gets (s^{-1} + s^{-1}X' + s^{-1}\overline{P})$ $mod$ $2^n$
    \ENDFOR
    \ENDFOR
\STATE \textbf{do}  $\overline{P} \gets \overline{Q}$, $s^{-1} \gets t^{-1}$, $X' \gets Y'$ repeat from 1 to 18 \textbf{for} $Y$
\STATE \textbf{return}  $\overline{P} = X$ and $\overline{Q} = Y$
\end{algorithmic}
\end{algorithm}
\subsubsection{logistic map}
\begin{figure}
    \centering
    \includegraphics[width=0.5\textwidth]{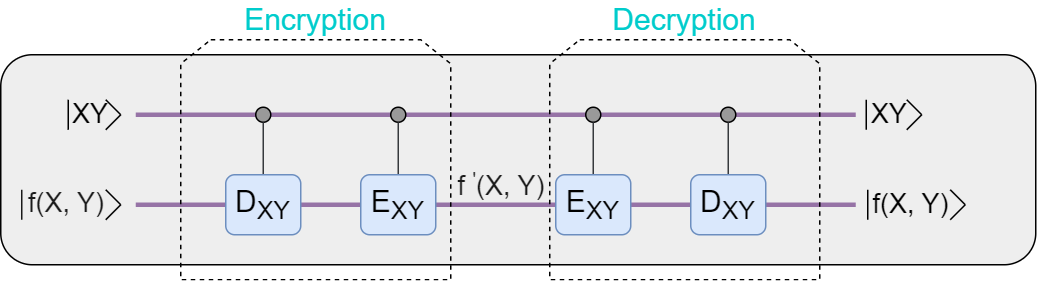}
    \caption{LM for encryption and decryption of pixel value guided by horizontal $X$ and vertical $Y$ position of pixels.}\label{NEQR_fig7}
\end{figure}

The inverse LM procedure is very simple with the output of Algo. \{\ref{NEQR_algo3}\}. This procedure operates the same functions $\phi_{YX}^1$ and $\phi_{YX}^2$ used in the encryption procedure (see Fig. \ref{NEQR_fig7}) as:
\begin{align}
         \Ket{\psi} &= \phi_{YX}^{1}\phi_{YX}^{2}\Ket{\psi_{L}}\nonumber\\
          &= \phi_{YX}^{1}\phi_{YX}^{2}\phi_{YX}^{2}\phi_{YX}^1\Ket{\psi}\nonumber\\
        &= \frac{1}{2^n}\sum_{Y, X=0}^{2^{2n} - 1} (D_{YX}E_{YX}E_{YX}D_{YX}\bigotimes_{i = 0}^{p - 1}\Ket{C_{YX}^{i}})\Ket{YX}\nonumber\\
        &= \frac{1}{2^n}\sum_{Y, X=0}^{2^{2n} - 1} \bigotimes_{i = 0}^{p - 1}\Ket{C_{YX}^i}\Ket{YX}.
\end{align}

%%%%%%%%%%%%%%%%%%%%%%%%%%%%%%%%%%%%%%%%%%%%%%%%%%%%%%%%%%%%%%%%%
\section{Evaluation}\label{sec4}
\begin{figure}
     \centering
     \begin{subfigure}[b]{0.237\textwidth}
        \centering
         \includegraphics[width=\linewidth]{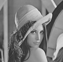}
         \caption{}
         %\label{NEQR_fig8a}
     \end{subfigure}
     \begin{subfigure}[b]{0.237\textwidth}
         \centering
         \includegraphics[width=\linewidth]{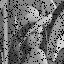}
         \caption{}
         %\label{NEQR_fig8b}
     \end{subfigure}
     \begin{subfigure}[b]{0.237\textwidth}
         \centering
         \includegraphics[width=\linewidth]{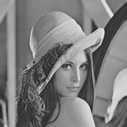}
         \caption{}
         %\label{NEQR_fig8c}
     \end{subfigure}
     \begin{subfigure}[b]{0.237\textwidth}
         \centering
         \includegraphics[width=\linewidth]{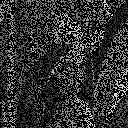}
         \caption{}
         %\label{NEQR_fig8d}
     \end{subfigure}
        \caption{The figure shows two original and two simulated images as per Fig.\ref{NEQR_fig1}: (a) The original Lena image of size $64 \times 64$. (b) The simulated result of the Lena image of size $64 \times 64$ with shots = 8192. (c) The original Lena image of size $128 \times 128$. (d) The simulated result of the Lena image of size $128 \times 128$ with shots = 8192. Increasing the number of shots/measurements enhances image quality by providing more probable states.}
        \label{NEQR_6}
\end{figure}

%\subsection{Image Reconstruction After Simulation}
Image reconstruction is performed using the pixel values and pixel positions obtained after the simulation of the quantum circuits (constructed according to Fig. \ref{NEQR_fig1}) in the Qiskit Qasm simulator, which resembles a real perfect quantum computer \cite{NEQR_Qiskit2019}. The results are presented in Fig. \ref{NEQR_6}.
\subsection{Correlation Coefficient Analysis}
\begin{figure*}
     \centering
     \begin{subfigure}[b]{0.3\textwidth}
         \centering
         \includegraphics[width=\textwidth]{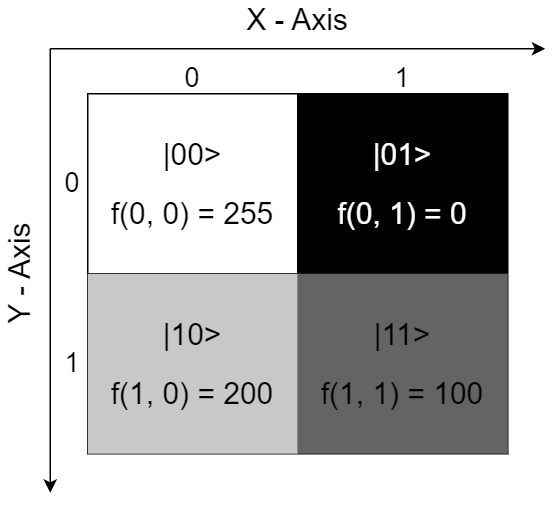}
         \caption{}
         \label{NEQR_fig8a}
     \end{subfigure}
     \hfill
     \begin{subfigure}[b]{0.3\textwidth}
         \centering
         \includegraphics[width=\textwidth]{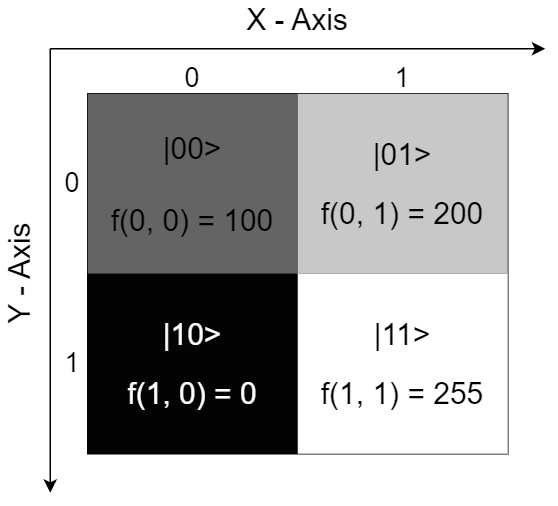}
         \caption{}
         \label{NEQR_fig8b}
     \end{subfigure}
     \hfill
     \begin{subfigure}[b]{0.3\textwidth}
         \centering
         \includegraphics[width=\textwidth]{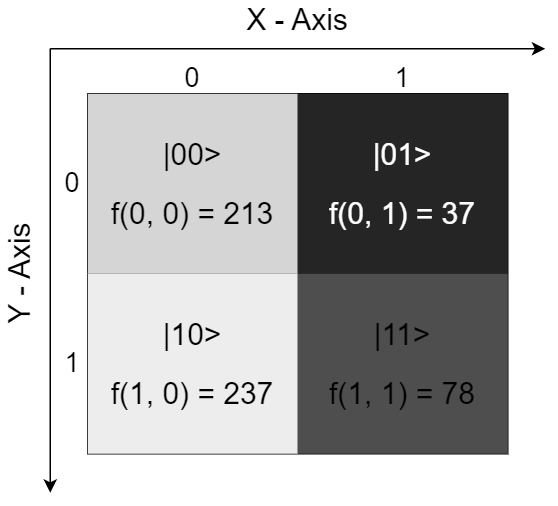}
         \caption{}
         \label{NEQR_fig8c}
     \end{subfigure}
        \caption{The figure shows three different images: (a) Original image for Eq. (\ref{image_eq}). (b) encrypted image using GAT. (c) Encrypted image using both GAT and LM. Here, the generalized affine encryption with $P$, $Q$, $s$, $t = 1$ and LM encryption with $L_0 = 0.5557924316949603$ and $\delta = 3.9816188727791215$.} is presented.
\end{figure*}
The correlation coefficient $r$ of two images is the analysis of how much these images are correlated to each other. The mathematical formula for the correlation coefficient is:
\begin{align}
    r = \frac{\sum (x_i - x_m) (y_i - y_m) }{\sqrt{\sum (x_i - x_m)^{2} \sum (y_i - y_m)^{2}}},\nonumber
\end{align}
where $x_i$ is the $i_{th}$ pixel intensity in the original image, $y_i$ is the $i_{th}$ pixel intensity in the encrypted image, $x_m$ \& $y_m$ are the mean pixels intensity of the original and encrypted image, respectively.

One can decide on the correlation based on the value of $r$ such that, 
\begin{itemize}
    \item $r = 1$: images are identical
    \item $r = 0$: images are uncorrelated
    \item $r = -1$: one image is the negative of the other
\end{itemize}
After using the values of each pixel intensity from Table \ref{pixel_value}, the value of the correlation coefficient $r$ turns out to be -1, which means that the encrypted image in Fig. \ref{NEQR_fig8c} is the negative image of the original image in Fig. \ref{NEQR_fig8a}.
\begin{table}
    \centering
    \begin{tabular}{c|c|c}
    \hline
        pixel position & plain & encrypted\\
        $\ket{YX}$ & image & image\\
        \hline
        $\Ket{00}$&255&213\\
        $\Ket{01}$&0&37\\
        $\Ket{10}$&200&237\\
         $\Ket{11}$&100&78\\
         \hline
         mean value & 139 & 141
    \end{tabular}
    \caption{Table shows the pixel intensity of plain and encrypted images corresponding to their position in the image.}
    \label{pixel_value}
\end{table}
\subsection{Encryption Analysis}
The effectiveness of the encryption procedure depends on the keyspace, key sensitivity, and its ability to generate different encrypted images for minor changes in the original image. 
\subsubsection{keyspace}
\begin{table}[]
    \centering
    \begin{tabular}{c|c|c}
    \hline
        image & $NPCR$ (in \%) & $UACI$ (in \%)\\
        \hline
        encrypted image&25&0.098\\
        (same parameter)&&\\
        encrypted image&100&28.04\\
        (different parameter)&&\\
    \end{tabular}
    \caption{Table shows the values of $NPCR$ and $UACI$. In the first raw, for the images in Fig. \ref{NEQR_fig8c} and in Fig. \ref{NEQR_figu9a}. In second raw, for the images in Fig. \ref{NEQR_fig8c} and in Fig. \ref{NEQR_figu9c}.}
    \label{neqr_Tab1}
\end{table}
keyspace is the storage space that consists of all the possible combinations of keys for an encrypted system such that one of them can decrypt the system perfectly. %Like - If someone encrypts data using a unique n-bit string of 0's and 1's, then to access the plain data without any knowledge of the key, someone has to try all possible combinations of n-bits, in total $2^n$ combinations or brute-force attack in real-time. 
The keyspace rises with the number of bits $(n)$, and can be represented as: \\
\begin{align}
keyspace (KS) \propto 2^n.\nonumber
\end{align}

For GAT, $n_P$, $n_Q$, $n_s$, $n_t$-qubit have been used to store the parameters $P$, $Q$, $s$, and $t$, respectively. Hence, the keyspace for GAT is $KS_A = 2^{n_P + n_Q + n_s + n_t}$. For LM, $n_{L_0}$ and $n_\delta$-qubit for $L_0$ and $\delta$, have been used. Hence, the keyspace is $KS_L = 2^{n_{L_0} + n_\delta}$. In total, the keyspace is $KS_T = KS_A + KS_L$.

To encrypt the image, one should select the encryption parameters $(P$, $Q$, $s$, $t$, $L_0$, $\delta)$ so that a brute-force attack (trying all possible combinations of the key) cannot break the encryption in the feasible amount of time. However, as discussed in \cite{NEQR_LiangQIP2016}, $L_0$ and $\delta$ are infinite decimals, so the keyspace becomes infinite. Therefore, the encrypted image resists brute-force attacks.

\subsubsection{Differential Analysis}
\begin{figure}
     \centering
     \begin{subfigure}[b]{0.237\textwidth}
        \centering
         \includegraphics[width=\linewidth]{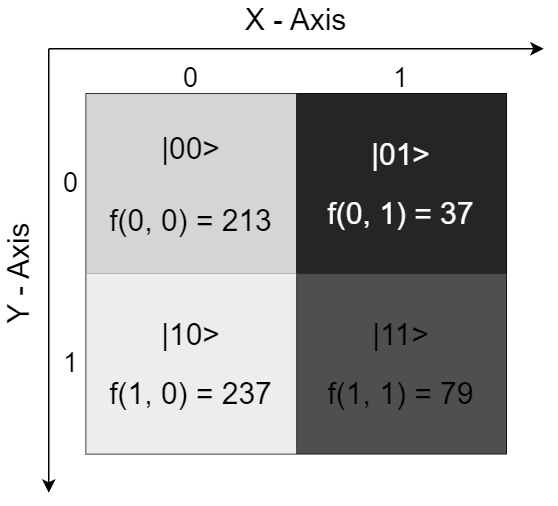}
         \caption{}
         \label{NEQR_figu9a}
     \end{subfigure}
     \begin{subfigure}[b]{0.237\textwidth}
         \centering
         \includegraphics[width=\linewidth]{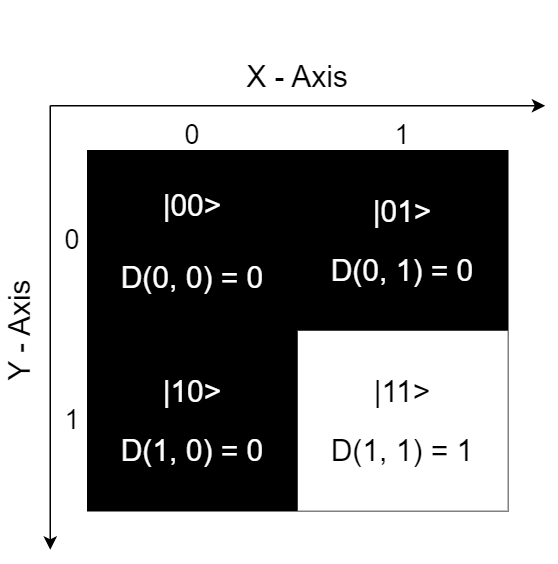}
         \caption{}
         \label{NEQR_figu9b}
     \end{subfigure}
     \begin{subfigure}[b]{0.237\textwidth}
         \centering
         \includegraphics[width=\linewidth]{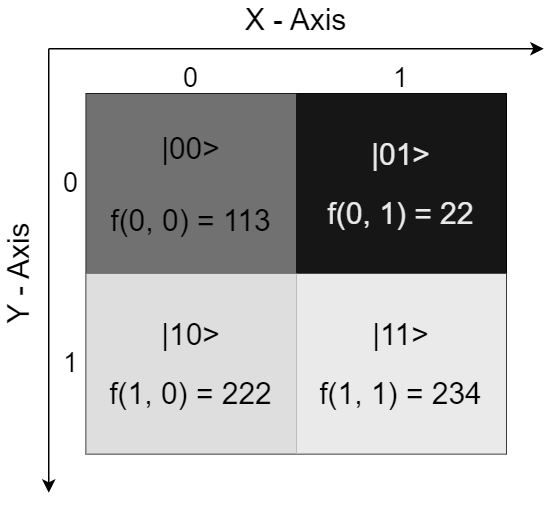}
         \caption{}
         \label{NEQR_figu9c}
     \end{subfigure}
     \begin{subfigure}[b]{0.237\textwidth}
         \centering
         \includegraphics[width=\linewidth]{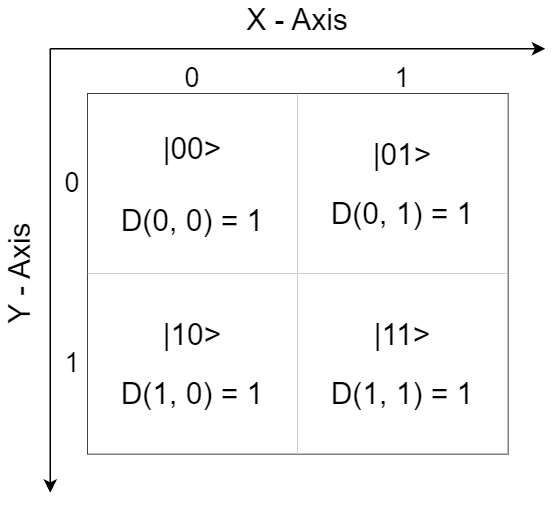}
         \caption{}
         \label{NEQR_figu9d}
     \end{subfigure}
        \caption{The figure shows two encrypted and two differential images: (a) is the encrypted image of Fig. \ref{NEQR_fig8a} with a change in the pixel value of the pixel at position (0, 0) from 255 to 254. (b) is the differential image of Fig. \ref{NEQR_figu9a} and Fig. \ref{NEQR_fig8c}. (c) is the encrypted image of Fig. \ref{NEQR_fig8a} with change in key $L_0$ from 0.5557924316949603 to 0.6, while the other keys are intact. (d) is a differential image of Fig. \ref{NEQR_figu9c} and Fig. \ref{NEQR_fig8c}.}
\end{figure}

Two differential analysis experiments were conducted to assess the sensitivity of the encryption procedure to changes in the encryption key and the original image. In the first experiment, the value of the pixel at position (0, 0) is changed from 255 to 254, producing the encrypted image shown in Fig. \ref{NEQR_figu9a}. Fig. \ref{NEQR_figu9b} shows the differential image between the image of the original encrypted image \ref{NEQR_fig8c} and the encrypted image after a change in the pixel value \ref{NEQR_figu9a}. The difference can be measured by the number of pixels change rate $(NPCR)$ and unified average changing intensity $(UACI)$ as follows:    
\begin{align}
    NPCR &= \frac{1}{W\times H}\left[\sum_{i, j} D(i, j)\right] \times 100\%,\nonumber\\
    UACI &= \frac{1}{W\times H}\left[\frac{\sum_{i, j}|C(i, j) - C'(i, j)|}{255}\right] \times 100\%,\nonumber
\end{align}
where $W$ and $H$ represent the width and height of the image, respectively. $D(i, j)$ is determine as,
\begin{align}
    D(i, j) = \left\{ \begin{array}{rcl}
1 & \mbox{for}& C(i, j)=C'(i, j), \\ 
0 & \mbox{for}& C(i, j)\neq C'(i, j),
\end{array}\right.\nonumber
\end{align}
where $C(i, j)$ and $C'(i, j)$ are the pixel values at the coordinate $(i, j)$ of the images in Fig. \ref{NEQR_fig8c} and in Fig. \ref{NEQR_figu9a}, respectively.

The second experiment consists of altering the value of the key parameter $L_0$ from 0.5557924316949603 to 0.6 while keeping all the other keys unchanged. The resulting encrypted image is depicted in Fig. \ref{NEQR_figu9c}. Then, the difference between the modified and original encrypted images is calculated. Finally, a differential image, illustrated in Fig. \ref{NEQR_figu9d}, is constructed that demonstrates that the new encrypted image is entirely different from the previous encrypted image \ref{NEQR_fig8c}. Thus, even a slight alteration in $L_0$ leads to generating a completely different image, highlighting the sensitivity of the encryption procedure to the key parameters.
% The second experiment is to change the key $L_0$ from 0.5557924316949603 to 0.6 while keeping the other keys constant. The encrypted image using $L_0 = 0.6$ is shown in Fig. \ref{NEQR_fig8c}. The differential image shown in Fig. \ref{NEQR_figu9d} is constructed with the original encrypted image \ref{NEQR_fig8c} and the new encrypted image. According to the differential image, it is apparent that the encryption procedure generates a completely different image if one of the keys is changed.

The $NPCR$s and $UACI$s are listed in Table \ref{neqr_Tab1}, which shows that our encryption method is not too sensitive to plain text, while it is highly sensitive to encryption keys (a slight change in encryption key results in a completely different encrypted image).
\subsubsection{Pear-signal-to-noise-ratio Analysis}
The peak-signal-to-noise-ratio $(PSNR)$ is a popular metric to measure the fidelity between two images. Here, the original image from Fig. \ref{NEQR_fig8a} and the encrypted image from Fig. \ref{NEQR_fig8c} are considered \cite{NEQR_Miyake2016, NEQR_Naseri2017, NEQR_Wang2004}. The higher the $PSNR$, the lower the quality of encryption. The $PSNR$ is defined as follows:
\begin{align}
    PSNR = 20log_{10}\left(\frac{MAX_{I}}{\sqrt{MSE}}\right),
\end{align}
where $MAX_{I}$ is the maximum possible pixel value of original image and $MSE$ is the mean squared error, which for two $m\times n$ monochromatic images is defined as \cite{NEQR_Sara2019},
\begin{align}
    MSE = \frac{1}{mn} \sum_{i=0}^{m - 1}\sum_{j=0}^{n - 1}[I(i, j) -K(i, j)],
\end{align}
where the $I(i, j)$ and $K(i, j)$ represents the pixel value of $i^{th}$ raw \& $j^{th}$ column of the original and encrypted image, respectively. The higher the value of $MSE$, the better the encryption procedure. The error is the quantity degradation by which the pixel value of the original image changes to form the encrypted image. The value of $MSE$ turns out to be 25 based on the images from Fig. \ref{NEQR_fig8a} and Fig. \ref{NEQR_fig8c}, and the $MAX_{I}$ is 255, so the $PSNR$ value is 34.1514 dB, which is acceptable because for 8-bit data the $PSNR$ value varies from 30 db to 50 db \cite{NEQR_Sara2019,NEQR_Deshpande2018}. 
\subsection{Histogram Analysis}
The analysis of the histogram includes image reconstruction using the information obtained from the histogram. In the histogram, the order of the state is represented as $\Ket{q_{9} \dots q_{2}q_{1}q_{0}} \rightarrow \Ket{C_{0} \dots C_{7}YX}$. By converting this binary state to decimal format, both the pixel location $(YX)$ and its corresponding pixel value $(f(Y, X))$ can be re-calculated. 

\subsection{Complexity analysis}
To assess the complexity of the quantum circuit, the following two key metrics are used: 1) quantum cost: the number of elementary quantum gates used in the quantum circuit and 2) time complexity: It denotes the depth or the number of time steps in the execution of the quantum circuit (where the execution of all basis operations in one step is taken as 1 unit). 
% \begin{table}[]
%     \centering
% \begin{tabular}{c|c|c|c}
% \hline
%      Method&color& Complexity & Retrieval  \\
%      &encoding& & \\
%      \hline
%      FRQI&grayscale&$\mathcal{O}(2^{4n})$&Probabilistic\\
%      MCQI&RGB&$\mathcal{O}(2^{4n})$&Probabilistic\\
%      NEQR&grayscale&$\mathcal{O}(qn.2^{2n})$&Deterministic\\
%      CQIR&grayscale&$\mathcal{O}(log_{2}L.n.2^{2n})$&Deterministic\\
%      QUALPI&grayscale&$\mathcal{O}(q(m+n).2^{m+n})$&Deterministic\\
%      SQR&infrared&$\mathcal{O}(2^{2n})$&Probabilistic\\
%      QSMC\&QSNC&RGB&$\mathcal{O}(2^{4n})$&Probabilistic\\
%      NAQSS&grayscale/RGB&$\mathcal{O}(log2^{n}.2^{2n})$&Probabilistic\\
%      \hline
% \end{tabular}
% \caption{....}
% \label{complexity_compair}
% \end{table}
\subsubsection{Quantum cost}
In our case, the part of the circuit that affects the quantum cost is the encoding of the pixel values according to the positions of the pixels because this part includes several $\mathbb{C}^{2n}$ (NOT gates). These gates are decomposed into $2(2n - 1)$ Toffoli gates and a c-NOT gate with $2n - 1$ ancillary qubit, where $\mathbb{C}^{2n}$ represents the $2n$ control qubits. In total, the decomposition of $\mathbb{C}^{2n}$ NOT gate into basis single and two-qubit gates will increase the cost enormously.

This increase in cost can be minimized with the help of the Espresso algorithm. The algorithm takes the control qubits information and returns a circuit that performs the same control task but with fewer control qubits. For example, the NEQR circuit in Fig. \ref{NEQR_fig10a}, which has 14 Toffoli gates is optimized and reduced to the circuit in Fig. \ref{NEQR_fig10b}, which has only 8 Toffoli gates by applying the Espresso algorithm. The quantum cost for the quantum circuit shown in Fig. \ref{NEQR_fig10a} is 234, which reduces to 154 after the application of the Espresso algorithm. To further optimize the quantum cost, the technique demonstrated in Fig. \ref{NEQR_fig10c} has been deployed, resulting in the conversion of our circuit to Fig. \ref{NEQR_fig10d} with a quantum cost of 91. This reduction in quantum cost is considerable; as with the Espresso algorithm and the presented technique, the total cost of the NEQR quantum circuit in Fig. \ref{NEQR_fig10a} decreases to 38.89\%, compared to 65.81\% without employing the technique depicted in Fig. \ref{NEQR_fig10c}.

%Which is 65.81\% without adding the presented technique shown in Fig. \ref{NEQR_fig10c}.

\subsubsection{Time complexity}
As discussed in \cite{NEQR_Zhang2013a}, the time complexity of the NEQR circuit is not greater than $\mathcal{O}(qn2^{2n})$. Similarly, in \cite{NEQR_LiangQIP2016}, it was shown that the time complexity of the encryption procedure is $\mathcal{O}(n)$, which is the same for the decryption procedure.

\begin{figure*}
\centering
        \begin{subfigure}[]{0.45\textwidth}
                \centering
                \includegraphics[height=0.55\textwidth, width=1.1\linewidth]{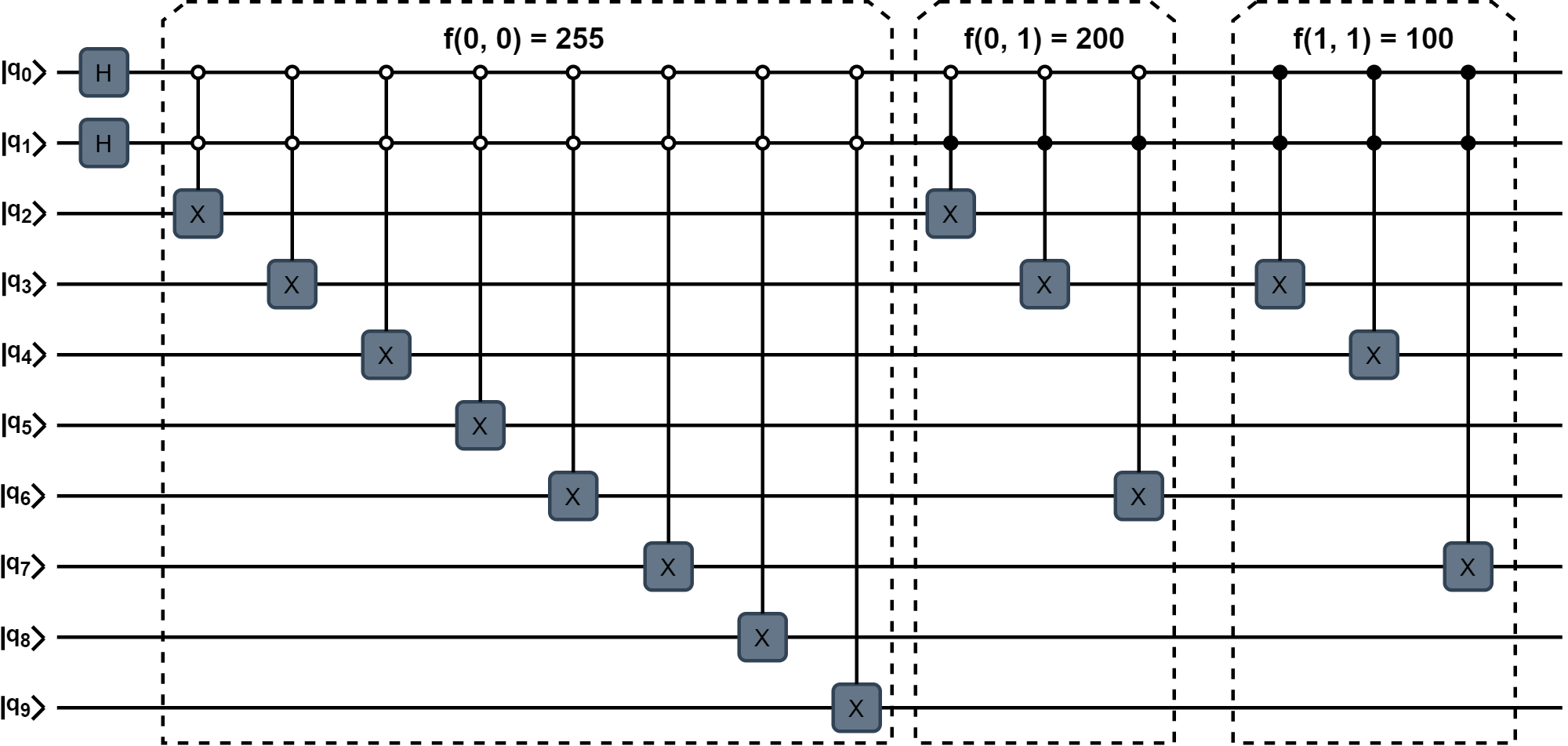}
                \caption{}\label{NEQR_fig10a}
        \end{subfigure}\hfill
        \begin{subfigure}[]{0.45\textwidth}
                \centering
                \includegraphics[width=\textwidth]{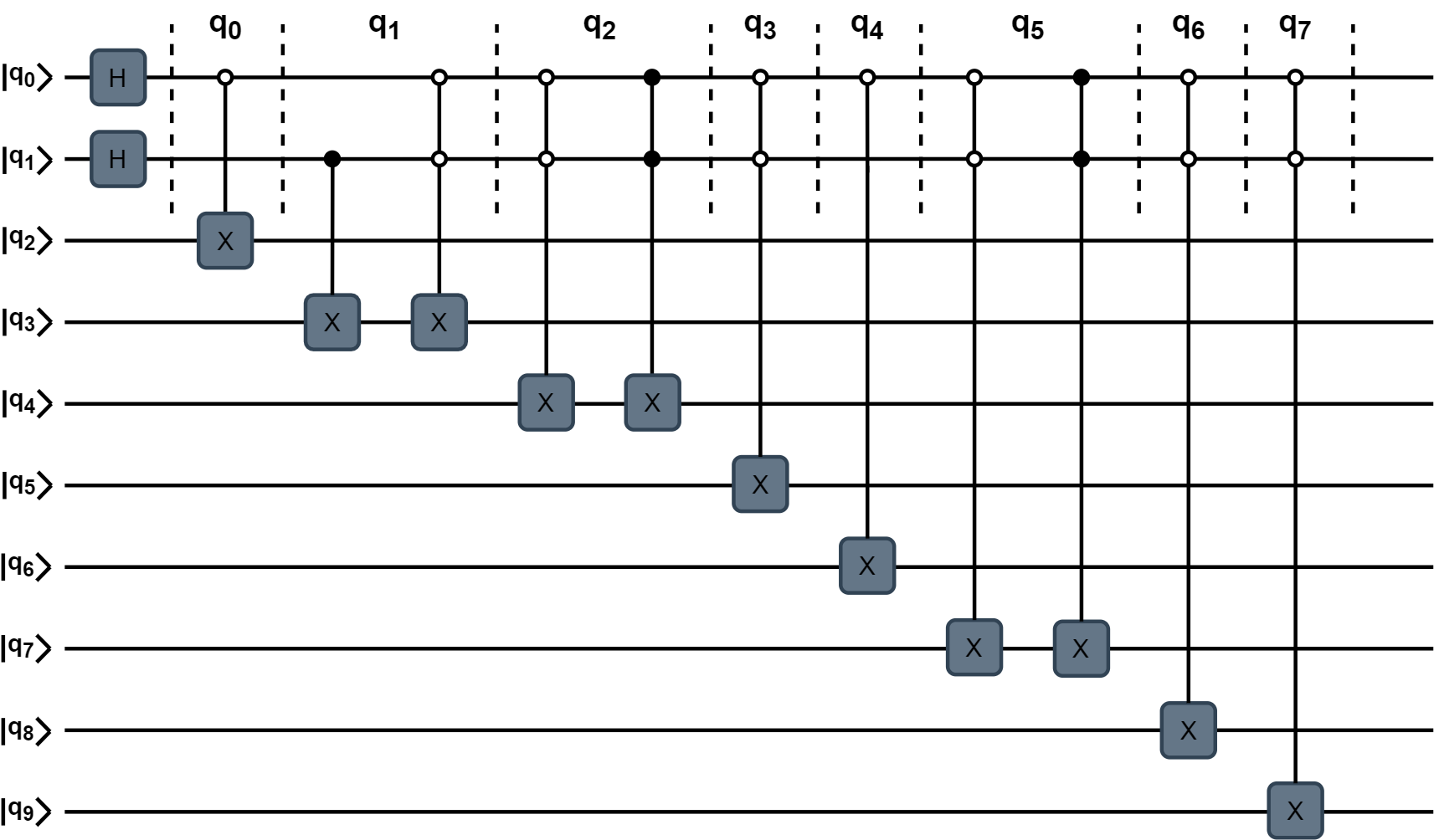}
                \caption{}\label{NEQR_fig10b}
        \end{subfigure}
        \begin{subfigure}[]{0.45\textwidth}
                \includegraphics[height=0.6\textwidth, width=1.05\linewidth]{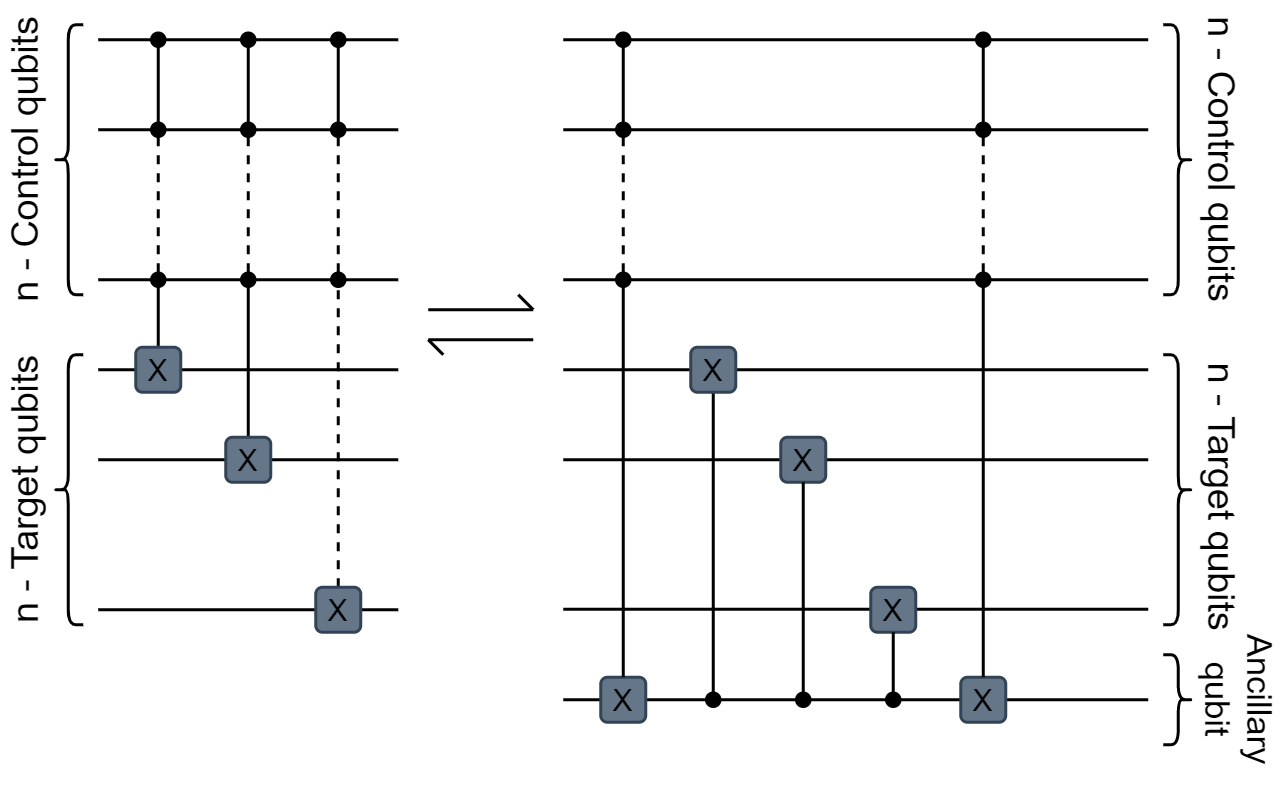}
                \caption{}\label{NEQR_fig10c}
        \end{subfigure}\hfill
        \begin{subfigure}[]{0.45\textwidth}
                \includegraphics[height=0.62\textwidth,width=\linewidth]{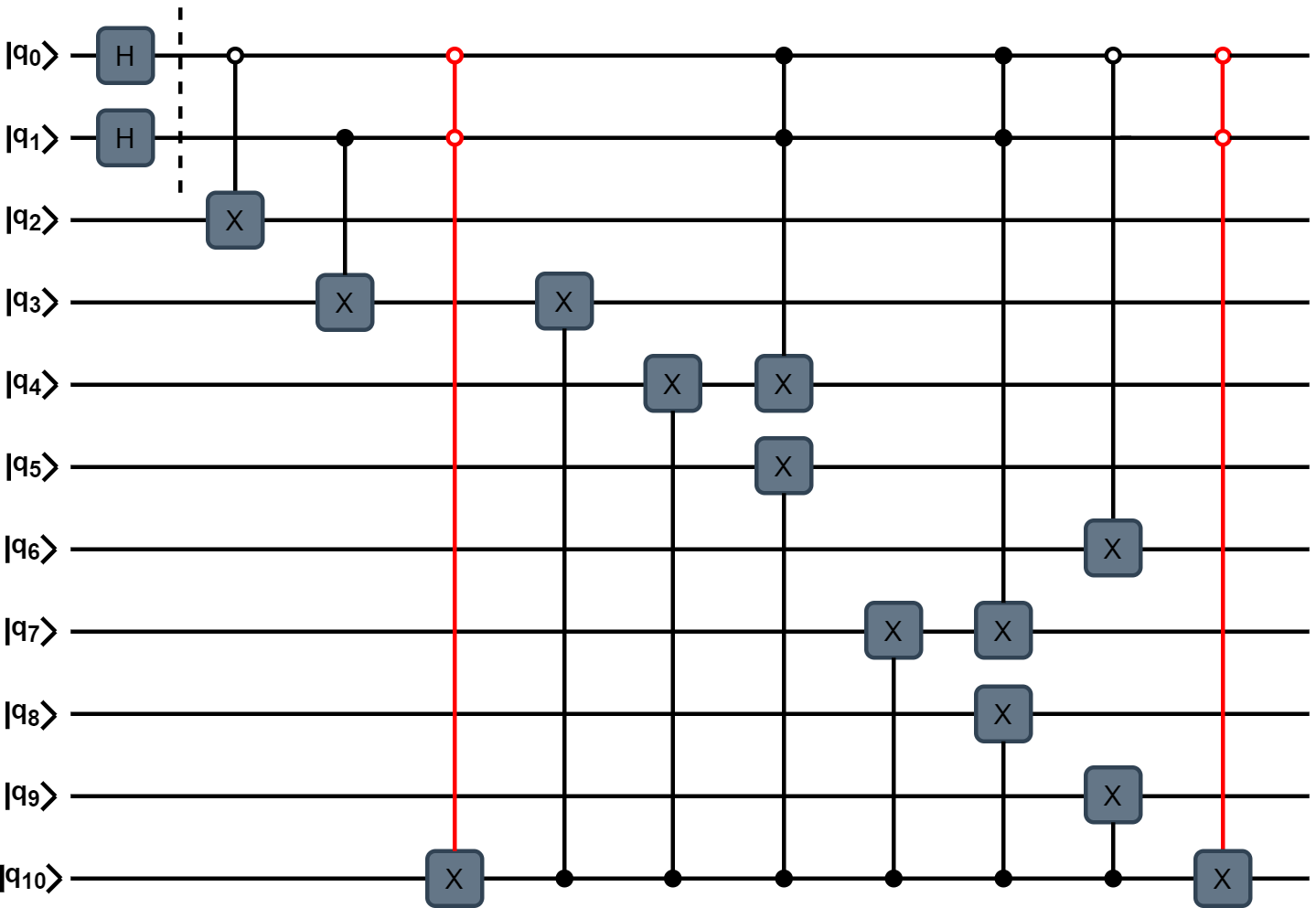}
                \caption{}\label{NEQR_fig10d}
        \end{subfigure}%
        \caption{(a) Quantum circuit for classical image shown in Fig. \ref{NEQR_fig8a}, according to the NEQR quantum model. Note that the position coordinates of the pixel are encoded into the circuit as $\Ket{YX} \rightarrow \Ket{{q_1}{q_0}}$ and the pixel values as $\Ket{C_{YX}^{0}C_{YX}^{1} \dots C_{YX}^{7}} \rightarrow \ket{q^{9}q^{8} \dots q^{2}}$. (b) The quantum circuit represents the circuit shown in Fig. \ref{NEQR_fig10a} after the reduction in the controlled information for the pixel values using the Espresso algorithm. (c) The figure shows the quantum circuit to reduce the number of Toffoli gates with the help of one ancillary qubit. (d) The figure shows the quantum circuit after the application of the Espresso algorithm and the technique presented in Fig.\ref{NEQR_fig10c}. The Toffoli gate represented by red is the one whose information is saved in the ancillary qubit.}
\end{figure*}

The quantitative analysis of the time complexity of the NEQR circuit, after encryption and decryption, is presented in Table \ref{complexity}. Additionally, after including the technique shown in Fig. \ref{NEQR_fig10c}, the time complexity is optimized significantly with the expense of adding one ancillary qubit. From Table \ref{complexity}, it is evident that the time complexity is reduced to 35.38\%,  a substantial improvement compared to 63.26\% without employing the technique shown in Fig. \ref{NEQR_fig10c}. 

%which was 63.26\%, without using the presented technique shown in Fig. \ref{NEQR_fig10c}.
\begin{table}[]
    \centering
\begin{tabular}{c|c|c|c}
\hline
     &Circuit& Quantum & Time  \\
     && cost& complexity\\
     \hline
     &NEQR & 234 & 147\\
     Normal&Encryption& 666 & 513 \\
     &Decryption& 1126 & 887\\
     \hline
     &NEQR& 154&93\\
     with Espresso&Encryption& 374&247\\
     &Decryption&597 &403\\
     \hline
     &NEQR& 91 &52\\
     including presented &Encryption&216 &134\\
     technique &Decryption& 342&215
\end{tabular}
\caption{Computational Complexities of the NEQR, NEQR with encryption, and NEQR with encryption and decryption quantum circuits for image \ref{NEQR_fig8a}. The calculation is done using the qiskit transpiler with optimization level $=$ 3 and basis gate set =  [cx, id, rz, sx, x], which is the most common basis gate set for IBM's quantum hardware. In the 2nd raw the values are calculated without any circuit compression methods. In 3rd raw the values are calculated with the Espresso algorithm compression method and in the 4th raw the values are calculated with the Espresso algorithm \& with the presented compression method shown in Fig. \ref{NEQR_fig10c}. The unit of quantum cost is the number of gates and time complexity is the time steps taken to operate all the possible quantum gates that can be simulated simultaneously.}
\label{complexity}
\end{table}
\subsection{Noise Analysis}
Current small-scale quantum computers are inherently noisy, necessitating consideration of noise effects for realistic simulations. In this section, we adopt six different types of noisy environments, namely amplitude-damping, phase-damping, bit-flip, phase-flip, bit-phase-flip, and depolarizing. These noisy channels are characterized by Kraus operators \cite{NEQR_Kraus1983}. To observe the effect of a particular noise on a circuit, the corresponding Kraus operator should affect each qubit of the circuit or the qubits that will have a noisy impact. This noise transforms the pure quantum state $(\Ket{\psi})$ to a mixed state, whose properties are best described by the density matrix $(\rho = \Ket{\psi}\bra{\psi})$. The effect of noise on the pure state $\psi$ is formulated as:
\begin{align}
    \epsilon^r(\rho) = \sum_m(\otimes_{i = 0}^n K^{rq_i}_m) \rho (\otimes_{i = 0}^n K^{rq_{i}\dagger}_m),\nonumber
\end{align}
where $K_{m}$ represents the Kraus operators, $r$ signifies the type of noisy environment, $q_i$ denotes the qubit on which noise acts, and $n$ is the number of qubits. The effect of noise on the real quantum state $\ket{\psi}$ can be quantified by calculating the fidelity. Fidelity is a measure of the closeness between two quantum states and is given by:
\begin{align}
    F(\rho, \epsilon(\rho)) = \left[Tr\sqrt{\sqrt{\rho}\epsilon(\rho)\sqrt{\rho}}\right]^2,
\label{fidelity}
\end{align}
where $\rho = \Ket{\psi}\bra{\psi}$ is the pure state, $\epsilon(\rho)$ is the noisy state, and $F(\rho, \epsilon(\rho)) \in [0, 1]$, where 1 means both states are identical, and 0 means the states are entirely different. 

Note that all noisy simulations are done only for the NEQR circuit (excluding the rest of the circuit) because of resource limitations. This is due to the dimension of the density matrix becoming $2^{15} \times 2^{15}$, which requires a substantial amount of memory and high computational power.   

\subsubsection{Amplitude-Damping Noise}
Amplitude damping describes the phenomenon where a qubit transitions from its excited state $\ket{1}$ to the ground state $\ket{0}$, signifying energy dissipation within the quantum system \cite{NEQR_Nielsen2010}. This process is also known as the $T_1$ relaxation \cite{NEQR_Sawaya2016, NEQR_Rost2020} and is modeled by the amplitude-damping quantum channel defined by the following Kraus operators:
\begin{align}
    K_0^A &= \left[\begin{matrix}
        1 & 0\\
        0 & \sqrt{1-\gamma_A}
    \end{matrix}\right], \quad K_1^A = \left[\begin{matrix}
        0 & \sqrt{\gamma_A}\\
        0 & 0
    \end{matrix}\right],\nonumber
\end{align}
where $\gamma_A \in [0, 1]$ is the probability of energy dissipation of the quantum system. This noisy channel leaves the $\ket{0}$ state of the qubit unchanged but changes the state $\ket{1}$ to the state $\sqrt{\gamma_A}\ket{0} + \sqrt{1-\gamma_A}\ket{1}$. Consequently, the quantum state of the image, as defined in Eq. (\ref{image_eq}), undergoes a transition to a mixed state, as depicted in Eq. (\ref{amp_noise_state}).

Fig. \ref{NEQR_fig11} clearly shows the effect of change in the probability of energy dissipation to the fidelity. It can also be observed from the fidelity diagram that the fidelity decreases as the dissipation probability goes from 0 to 1.
\begin{align}
    \epsilon^A(\rho) = &\otimes_{i = 0}^9 K^{Aq_i}_0 \rho \otimes_{i = 0}^9 K^{Aq_{i}\dagger}_0 \nonumber\\ &+ \otimes_{i = 0}^9 K^{Aq_i}_1 \rho_I \otimes_{i = 0}^9 K^{Aq_{i}\dagger}_1.\label{amp_noise_state}
\end{align}
\subsubsection{Phase-Damping Noise}
Phase damping refers to the perturbation observed in the off-diagonal elements of the density matrix, leading to a loss in relative phase information within the quantum state \cite{NEQR_Nielsen2010}. This effect is closely related to the qubit's dephasing time $(T_2)$ \cite{NEQR_Sawaya2016, NEQR_Rost2020} and is modeled through phase-damping quantum channels, characterized by Kraus operators:
\begin{align}
 K_0^P &= \left[\begin{matrix}
        \sqrt{1-\gamma_P} & 0\\
        0 & \sqrt{1-\gamma_P}
    \end{matrix}\right], \quad K_1^P = \left[\begin{matrix}
        \sqrt{\gamma_P} & 0\\
        0 & 0
    \end{matrix}\right], \nonumber\\
    K_2^P &= \left[\begin{matrix}
        0 & 0\\
        0 & \sqrt{\gamma_P}
        \end{matrix}\right],
        \label{pd_eq}
\end{align}
where $\gamma_P \in [0, 1]$ represents the probability of loosing relative Phase information. The effect of the Phase damping channel on a single qubit state, based on the Kraus operators from Eq. (\ref{pd_eq}), is as follows:
\begin{align}
    \ket{0}\rightarrow(\sqrt{\gamma_P}+\sqrt{1-\gamma_P})\ket{0},\nonumber\\
    \ket{1}\rightarrow(\sqrt{\gamma_P}+\sqrt{1-\gamma_P})\ket{1}.\nonumber
\end{align}
After passing through this noise channel the original quantum state of the image, as outlined in Eq. (\ref{image_eq}), transitions into a mixed state represented by Eq. (\ref{pd_noise_state}). The fidelity $F^P$ of the output state is calculated using Eq. (\ref{fidelity}).
The fidelity variations, corresponding to increasing rates of phase-damping noise, are shown in Fig. \ref{NEQR_fig11}.
\begin{align}
    \epsilon^P(\rho) = &\otimes_{i = 0}^{10} K^{Pq_i}_0 \rho \otimes_{i = 0}^{10} K^{Pq_{i}\dagger}_0 \nonumber\\&+ \otimes_{i = 0}^{10} K^{Pq_i}_1 \rho \otimes_{i = 0}^{10} K^{Pq_{i}\dagger}_1.\label{pd_noise_state}
\end{align}
\subsubsection{Bit-Flip Noise}
The bit-flip noisy channel probabilistically flips the state of a quantum bit with a probability parameter $\gamma_B \in [0, 1]$. The Pauli-X gate is known as a bit-flip gate with $\gamma_B = 1$ \cite{NEQR_Nielsen2010}. This noisy quantum channel is modeled in a quantum circuit using the following Kraus operators:
\begin{align}
    K_0^B &= \left[\begin{matrix}
        \sqrt{1-\gamma_B} & 0\\
        0 & \sqrt{1-\gamma_B}
    \end{matrix}\right], \quad K_1^B = \left[\begin{matrix}
        0 & \sqrt{\gamma_B}\\
        \sqrt{\gamma_B} & 0
    \end{matrix}\right].\nonumber
\end{align}
When the quantum state, as described in Eq. (\ref{image_eq}), passes through the bit-flip noisy channel, it transforms into a mixed quantum state represented by the density matrix in Eq. (\ref{bit_noise_state}). Notably, when a single qubit state undergoes this noisy channel, the effects manifest as follows,
\begin{align}
    \ket{0}\rightarrow\sqrt{1-\gamma_B}\ket{0} + \sqrt{\gamma_B}\ket{1},\nonumber\\
    \ket{1}\rightarrow\sqrt{\gamma_B}\ket{0} + \sqrt{1-\gamma_B}\ket{1}.\nonumber
\end{align}
The fidelity of the resulting noisy quantum state is evaluated using Eq. (\ref{fidelity}), with Figure \ref{NEQR_fig11} illustrating the variation in fidelity across different bit-flip probabilities $\gamma_B$ ranging from 0 to 1.
\begin{align}
    \epsilon^B(\rho) = &\otimes_{i = 0}^9 K^{Bq_i}_0 \rho \otimes_{i = 0}^9 K^{Bq_{i}\dagger}_0 \nonumber\\&+ \otimes_{i = 0}^9 K^{Bq_i}_1 \rho \otimes_{i = 0}^9 K^{Bq_{i}\dagger}_1.\label{bit_noise_state}
\end{align}
\subsubsection{Phase-Flip Noise}
In the phase-flip noisy channel, the state of a quantum system is unaffected, but the relative phase of the quantum system flips probabilistically with a parameter $\gamma_W \in [0, 1]$. The Pauli-Z gate is recognized as a phase-flip gate with $\gamma_W = 1$ \cite{NEQR_Nielsen2010}. This noisy quantum channel is represented in a quantum circuit using the following Kraus operators:
\begin{align}
    K_0^W &= \left[\begin{matrix}
        \sqrt{1-\gamma_W} & 0\\
        0 & \sqrt{1-\gamma_W}
    \end{matrix}\right], \quad K_1^B = \left[\begin{matrix}
        \sqrt{\gamma_W} & 0\\
        0 & -\sqrt{\gamma_W}
    \end{matrix}\right].\nonumber
\end{align}
After transmission through this noisy channel, the quantum state becomes a mixed state, resulting in the loss of phase information from the original state. Specifically, when the quantum state described in Eq. (\ref{image_eq}) passes through a phase-flip noisy channel, it transforms into Eq. (\ref{pf_noise_state}). Similarly, the effects on a single qubit state passing through this noisy channel are as follows:
\begin{align}
    \ket{0}\rightarrow(\sqrt{\gamma_W}+\sqrt{1-\gamma_W})\ket{0},\nonumber\\
    \ket{1}\rightarrow(\sqrt{\gamma_W}+\sqrt{1-\gamma_W})\ket{1}.\nonumber
\end{align}
The fidelity of the Eq. (\ref{pf_noise_state}) with Eq. (\ref{image_eq}) is shown in Fig. \ref{NEQR_fig11} with all possible values of $\gamma_W$ from 0 to 1 and is calculated using Eq. (\ref{fidelity}).
\begin{align}
    \epsilon^W(\rho) = &\otimes_{i = 0}^9 K^{Wq_i}_0 \rho \otimes_{i = 0}^9 K^{Wq_{i}\dagger}_0 \nonumber\\&+ \otimes_{i = 0}^9 K^{Wq_i}_1 \rho \otimes_{i = 0}^9 K^{Wq_{i}\dagger}_1.\label{pf_noise_state}
\end{align}
\subsubsection{Bit-Phase-Flip Noise}
In the bit-phase-flip noisy channel, a quantum state undergoes concurrent bit-flip and phase-shift operations with a probability $\gamma_F \in [0, 1]$. The Pauli-Y gate is a bit-phase-flip gate with $\gamma_F = 1$ \cite{NEQR_Nielsen2010}. The Kraus operators to model this noisy channel in a quantum circuit are:
\begin{align}
    K_0^F &= \left[\begin{matrix}
        \sqrt{1-\gamma_F} & 0\\
        0 & \sqrt{1-\gamma_F}
    \end{matrix}\right], \quad K_1^F = \left[\begin{matrix}
        0&-\iota\sqrt{\gamma_W}\\
        \iota\sqrt{\gamma_W}&0
    \end{matrix}\right].\nonumber
\end{align}
The mixed state after passing through this noisy channel is written in Eq. (\ref{bf_noise_state}) and when a single qubit state passes through this noisy channel, the effects are,
\begin{align}
    \ket{0}\rightarrow\sqrt{1-\gamma_F}\ket{0}+\iota\sqrt{\gamma_F}\ket{1},\nonumber\\
    \ket{1}\rightarrow-\iota\sqrt{\gamma_F}\ket{0}+\sqrt{1-\gamma_F})\ket{1}.\nonumber
\end{align}
The variation in the fidelity as per the flipping probability $\gamma_F$ is shown in Fig. \ref{NEQR_fig11}, which is calculated by using Eq. (\ref{fidelity}).
\begin{align}
    \epsilon^F(\rho) = &\otimes_{i = 0}^9 K^{Fq_i}_0 \rho \otimes_{i = 0}^9 K^{Fq_{i}\dagger}_0 \nonumber\\&+ \otimes_{i = 0}^9 K^{Fq_i}_1 \rho \otimes_{i = 0}^9 K^{Fq_{i}\dagger}_1.\label{bf_noise_state}
\end{align}
\subsubsection{Depolarization Noise}
\begin{figure}
    \centering
    \includegraphics[width = 0.45\textwidth]{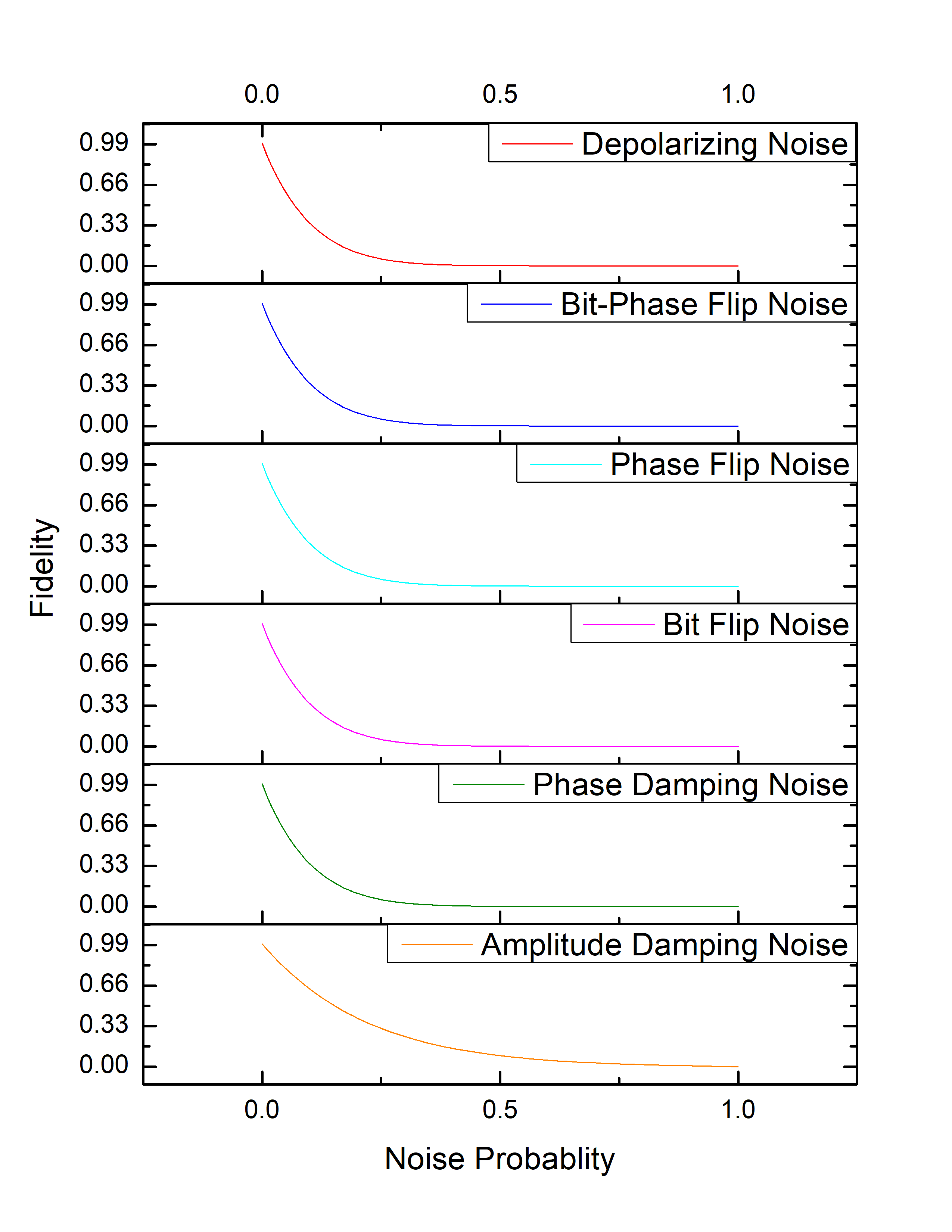}
    \caption{The plot shows the variation in the fidelity for the quantum state shown in Eq. (\ref{image_eq}) with the change in noise rate in six different types of noisy channels. Here, the fidelity variation comes out to be similar for all the noisy environments except the Amplitude damping noisy environment.}\label{NEQR_fig11}
\end{figure}

In Depolarizing noisy environment, the quantum system experiences depolarization with a probability $\gamma_D \in (0, 1)$. This noisy channel applies the Identity, Pauli-X, Pauli-Y, and Pauli-Z operators on the quantum system with probabilities $1-\gamma_D$, $\gamma_D/3$, $\gamma_D/3$, and $\gamma_D/3$, respectively. The Kraus operators representing this depolarizing noisy channel in a quantum circuit are:
\begin{align}
   K_0^D &= \left[\begin{matrix}
        \sqrt{1-\gamma_D} & 0\\
        0 & \sqrt{1-\gamma_D}
    \end{matrix}\right], \quad K_1^D = \left[\begin{matrix}
        0&\sqrt{\frac{\gamma_D}{3}}\\
        \sqrt{\frac{\gamma_D}{3}}&0
    \end{matrix}\right],\nonumber\\
    K_2^D &= \left[\begin{matrix}
        0&-\iota\sqrt{\frac{\gamma_D}{3}}\\
    \iota\sqrt{\frac{\gamma_D}{3}}&0\end{matrix}\right], \quad K_3^D = \left[\begin{matrix}
        \sqrt{\frac{\gamma_D}{3}}&0\\
        0&\sqrt{\frac{\gamma_D}{3}}
    \end{matrix}\right].\nonumber
\end{align}
Upon transmission through this noisy channel, the original quantum state described in Eq. (\ref{image_eq}) transforms into the mixed state detailed in Eq. (\ref{de_noise_state}). For a single qubit state passing through this noisy channel, the effects are as follows:
\begin{align}
    \ket{0}\rightarrow\left(\sqrt{1-\gamma_D} + \sqrt{\frac{\gamma_D}{3}}\right)\ket{0} + (1+\iota)\sqrt{\frac{\gamma_D}{3}}\ket{1}\nonumber\\
    \ket{1}\rightarrow(1-\iota)\sqrt{\frac{\gamma_D}{3}}\ket{1} + \left(\sqrt{1-\gamma_D} + \sqrt{\frac{\gamma_D}{3}}\right)\ket{1}\nonumber 
\end{align}
The fidelity, calculated using Eq. (\ref{fidelity}), shows variations as the depolarizing probability $\gamma_D$ changes, as illustrated in Fig. \ref{NEQR_fig11}.
\begin{align}
    \epsilon^D(\rho) = &\otimes_{i = 0}^{9} K^{Dq_{i}}_{0} \rho \otimes_{i = 0}^{9} K^{Dq_{i}\dagger}_{0} \nonumber\\&+ \otimes_{i = 0}^9 K^{Dq_i}_1 \rho \otimes_{i = 0}^9 K^{Dq_{i}\dagger}_1.\label{de_noise_state}
\end{align}

%%%%%%%%%%%%%%%%%%%%%%%%%%%%%%%%%%%%%%%%%%%%%%%%%%%%%%%%%%%%%%%%%
\section{Conclusion}\label{sec5}
In this paper, we presented an efficient implementation scheme of a novel enhanced quantum representation (NEQR) for a $2^n \times 2^n$ image, along with quantum image encryption and its decryption procedure on an IBM quantum computer. These algorithms are part of the fittest techniques to represent and encrypt a digital image because of their measurement accuracy and chaotic output with high keyspace and sensitivity. Furthermore, we have shown how to realize each of these procedures on a quantum computer and provided general quantum circuits and algorithms for them. The multi-control-NOT gates are essential to encode the pixel values, whose decomposition into basis one and two-qubit gates increases the circuit complexity as n increases. However, we have provided a method combined with the image compression method to reduce the effect of decomposition on circuit complexity and successfully reduced it to approximately 50\% compared to Zhang et al. for a $2 \times 2$ test image on the cost of adding one more qubit to the circuit. Regardless, it is very effective for circuit complexity to make this trade-off between a little extra quantum storage and a significant reduction in circuit complexity. Analysis of the encryption procedure demonstrated that the scheme is sensitive to the encryption key and has enough keyspace to resist ordinary attacks and attacks similar to brute-force attacks. In conclusion, we showed that NEQR, combined with the discussed novel quantum image encryption method, forms a more efficient, secure, feasible, and accurately measurable quantum image processing system. In the future, we would like to demonstrate the implementation of more quantum image representation protocols with an encryption procedure on the quantum computer. This leads to a quantum image processing system with a lower circuit complexity, higher measurement accuracy, and susceptible encryption technique that can be realizable on a quantum computer.

\end{document}